\documentclass[aps,rmp,preprint,groupedaddress,floatfix]{revtex4-1}
\usepackage{graphicx,subfigure,url,natbib,amsmath,amsthm,amssymb,amsfonts, cancel, setspace, ulem, sidecap}

\usepackage[title, titletoc, toc]{appendix}

\usepackage{genetics}

\allowdisplaybreaks

\begin{document}

\makeatletter
\newcommand{\manuallabel}[2]{\def\@currentlabel{#2}\label{#1}}
\makeatother

\title{The Dynamics of Genetic Draft in Rapidly Adapting Populations}
\author{Katya Kosheleva}
\author{Michael Desai}
\affiliation{\mbox{Department of Organismic and Evolutionary Biology, Department of Physics, and} \mbox{FAS Center for Systems Biology, Harvard University}}

\begin{abstract}

The accumulation of beneficial mutations on many competing genetic backgrounds in rapidly adapting populations has a striking impact on evolutionary dynamics. This effect, known as clonal interference, causes erratic fluctuations in the frequencies of observed mutations, randomizes the fixation times of successful mutations, and leaves distinct signatures on patterns of genetic variation. Here, we show how this form of `genetic draft' affects the forward-time dynamics of site frequencies in rapidly adapting asexual populations. We calculate the probability that mutations at individual sites shift in frequency over a characteristic timescale, extending Gillespie's original model of draft to the case where many strongly selected beneficial mutations segregate simultaneously. We then derive the sojourn time of mutant alleles, the expected fixation time of successful mutants, and the site frequency spectrum of beneficial and neutral mutations. We show how this form of draft affects inferences in the McDonald-Kreitman test, and how it relates to recent observations that some aspects of genetic diversity are described by the Bolthausen-Sznitman coalescent in the limit of very rapid adaptation. Finally, we describe how our method can be extended to model evolution on fitness landscapes that include some forms of epistasis, such as landscapes that are partitioned into two or more incompatible evolutionary trajectories.

\end{abstract}

\date{\today}

\maketitle

\vspace{0.5in}

Running Head: Genetic Draft in Rapidly Adapting Populations  \\


Corresponding Author:

Michael M. Desai

Departments of Organismic and Evolutionary Biology and of Physics

FAS Center for Systems Biology

Harvard University

435.20 Northwest Labs

52 Oxford Street

Cambridge, MA 02138

617-496-3613

mdesai@oeb.harvard.edu

\clearpage

\newpage

\section{Introduction}

The effects of linkage between beneficial mutations in altering evolutionary dynamics and the structures of genealogies in adapting populations has been recognized for nearly a century, particularly in the context of the evolutionary advantage of sex \citep{Muller1}. In both asexually reproducing organisms and in regions of low recombination in sexual organisms, the chance congregation of beneficial mutations on competing genetic backgrounds skews evolutionary dynamics. Because of this ``clonal interference'' effect, the success of a mutation depends not only on its fitness effect, but also on the quality of the genetic background in which it occurs and the fortune of the mutant's progeny in amassing more beneficial mutations  \citep{Smith1, Gerrish1, Gillespie1, Gillespie2, KimOrr} .

Recent work in experimental evolution has confirmed that clonal interference is widespread in large adapting laboratory microbial and viral populations \citep{Lang1, KaoSherlock, Miralles, DeVisser1, DeVisser2}. Several recent studies also suggest that classical ``hard'' selective sweeps may be rare in Drosophila \citep{Sella1, Karasov1} and humans \citep{Hernandez1,Pritchard1} implying that models that better account for linkage between sites need to be explored. As a result, in recent years there has been an influx of theoretical work describing the effects of clonal interference on the evolution of large populations (see \citet{Park1} for a recent review). 

\begin{figure}[h!]
\centering
	\subfigure
{		\includegraphics[width=9cm]{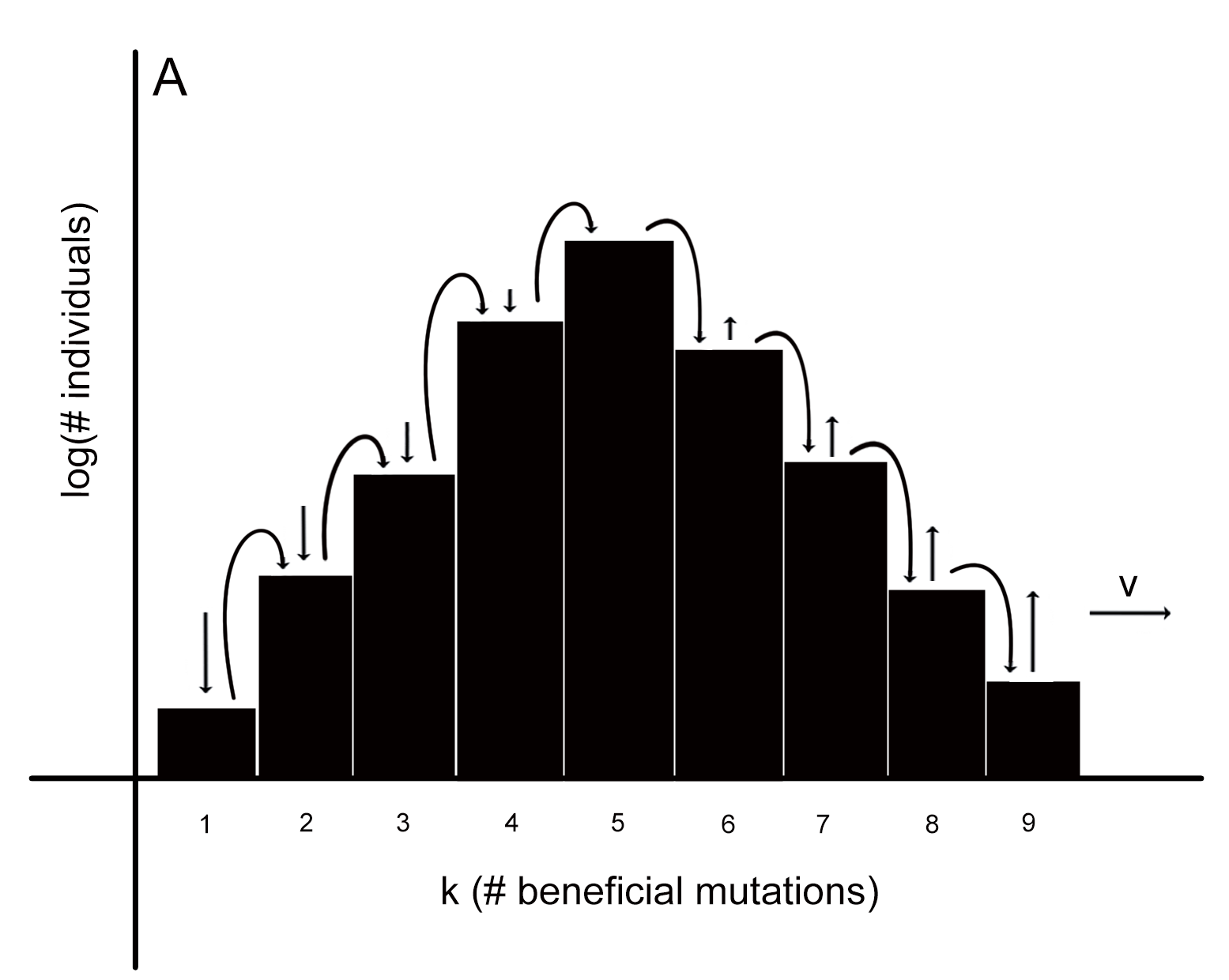}
}
%\hspace{1cm}
	\subfigure
{		\includegraphics[width=9cm]{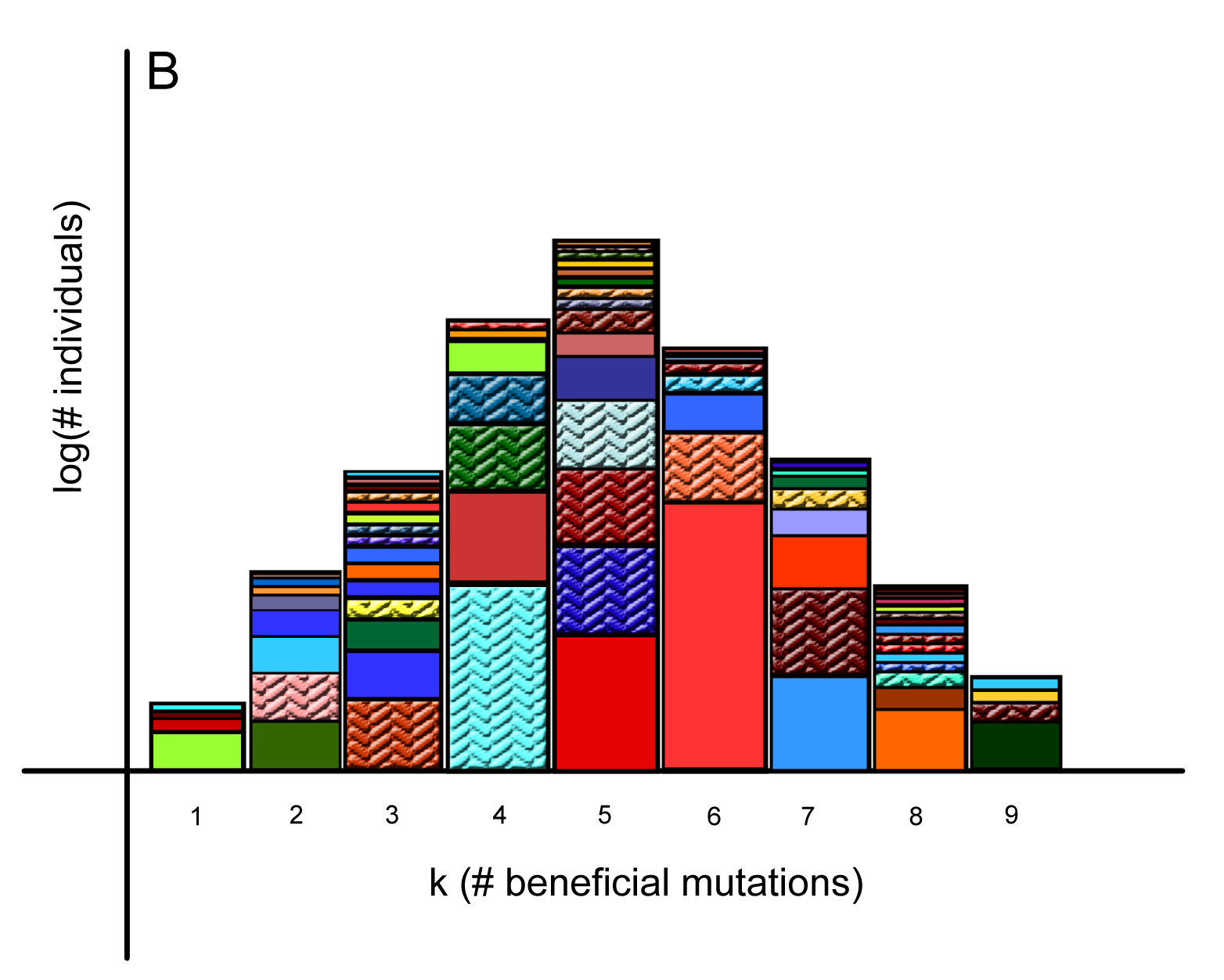}		
}
%\begin{center}
%\includegraphics[width=0.8\textwidth]{fitnessclassesfinished}
\caption{The profile of the fitness distribution at one instant in time. Each fitness class is indexed according to the number of beneficial mutations carried by individuals in that class, with each beneficial mutation conferring an identical fitness effect. $A$ illustrates the distribution of fitnesses without regards to the underlying haplotypes constituting each class.  Vertical arrows denote the effect of selection on each class (scaled by the strength of selection) and arrows between classes denote outgoing beneficial mutations (unscaled). By contrast, $B$ illustrates the same population, but divides each fitness class according to the contributions of its most successful haplotypes, represented by different colors within each class. The contribution in each class of a mutation originally arising in class 2 is also displayed, with haplotypes carrying the mutation represented by textured tiles.  \label{fitnessclasses}}
%\end{center}
\end{figure}

This work has provided a good understanding of evolutionary dynamics in the regime of \textit{rare} interference, where the number of strongly beneficial mutations segregating in a population is rarely more than two \citep{Gerrish1, Park2, Gillespie1, Gillespie2, Kim1}. However, in large populations many beneficial mutations can segregate simultaneously, and the population can maintain substantial variation in fitness. This decreases the importance of each mutant's intrinsic fitness effect relative to the quality of the genetic background on which it occurs. Long-term evolutionary dynamics in these populations are therefore driven primarily by the stochastic introduction of mutants at the high-fitness tip of the population's fitness distribution, and the fluctuation in the lineage sizes of these super-fit mutants when rare. Several models have been introduced to study evolution in these strong selection, strong mutation regimes \citep{DF, Rouzine1, Hallatschek1, Tsimring}. This work has successfully described the rate of adaptation and the variation in fitness within a population \citep{DF, Rouzine2, Park1}, and the fitness effects of fixed mutations \citep{FisherMagnumOpus, Neher3, Good1, Fogle}, while ignoring the specific mutations that underlie these population-wide quantities  (Figure \ref{fitnessclasses}A).

 In the present work, we use these earlier theoretical treatments as the basis for analyzing the evolutionary dynamics of individual mutations (i.e. their frequencies over time and their eventual fates). To do so, we study the forward-time dynamics of specific mutant lineages on the backdrop of the population's fitness distribution (Figure \ref{fitnessclasses}B). Our approach is complementary to recent work that analyzes diversity by considering the structure of genealogies in rapidly adapting populations, moving backwards in time from a sample of the present population \citep{DWF, Neher2}.
 % Our approach is complementary to recent work that analyzes diversity in rapidly adapting populations in a coalescent framework --- investigating the structure of genealogies through the ancestral history of a sample from the present population \citep{DWF, Neher2}.
  Our work also complements earlier analysis of related questions in facultatively sexual populations \citep{Neher1}, which neglect new mutations and focus instead on fluctuations in the frequencies of individual polymorphisms driven by recombination into higher or lower fitness backgrounds. By contrast, we focus on either asexual populations or on tightly linked genomic regions of sexual populations, where recombination can be neglected compared to selection and new mutations, and study instead the fluctuations in polymorphism frequencies driven by new mutations.

We begin by introducing our model and briefly summarizing earlier results that describe the dynamics of the population's fitness distribution. We then demonstrate that the growth of the high-fitness ``nose'' of this fitness distribution is dominated by a small number of successful, founding mutants. Since this high-fitness ``nose'' will eventually come to dominate the population, the long-term success of a given polymorphism is largely determined by its representation (or lack thereof) among this small class of stochastically fluctuating, high-fitness mutants. This allows us to model adaptation as a series of replacements of each fittest class by a new, fitter class over a typical replacement timescale.
We show how this leads to a distribution of transition probabilities describing how the frequency of each polymorphism changes in each stochastic jump from one fittest class to the next. This process bears some resemblance to several recent models of adaptation in populations with highly skewed offspring distributions \citep{Der, EldonWakeley}. However, whereas in these earlier models a jump in offspring frequency is assumed to be an explicit feature of the offspring distribution, in this work these jumps emerge organically from the dynamics of the underlying model.

We next use these transition probabilities to derive various diversity statistics, providing an alternative forward-time perspective that complements earlier structured coalescent approaches to these questions \citep{DWF}. We first calculate the site frequency spectrum of beneficial and neutral mutations, which has not yet been explicitly derived for this class of models. We then use our results to make predictive estimates regarding the fates of mutations in experiments, particularly on the sojourn time of these mutations and the time to fixation of a successful mutant. Finally, in the Discussion we describe a \textit{decay to neutrality} exhibited by mutations in these populations, comment on the relationship between our results and the Bolthausen-Sznitman coalescent, analyze the implications of our results for interpreting widely used tests for adaptation, and consider the extension of our model to evolution on more complex fitness landscapes.

\subsection{Heuristics}

To develop some intuition for the analysis to come, we observe that compared to the regime of successive selective sweeps, pervasive clonal interference leads to several interesting and somewhat unexpected consequences for the frequency trajectories of individual mutations. As noted above, the fate of any mutation in a rapidly adapting population is primarily determined by two factors: (1) the genetic background in which it occurs (or, more precisely, the net fitness of the resulting mutant) and (2) the success of its progeny in amassing additional beneficial mutations more quickly than competing backgrounds. 

When many beneficial mutations segregate simultaneously, the strictness of these two constraints requires that mutations with any non-negligible chance of fixing (or even rising to an appreciable frequency) must have been founded among the most fit individuals in the population. The vast majority of beneficial mutations are thus ``wasted'' on the bulk of the distribution, where the mutant's lineage is doomed to eventual extinction. As a result, frequencies of mutant alleles can be divided into two regimes. On the one hand, mutations at appreciable frequencies, which were almost certainly founded at the exponentially expanding high-fitness front of the fitness distribution, should exhibit site frequency spectra with some similarity to an exponentially growing population. Conversely, the majority of mutations (beneficial or otherwise) occur in the body of the population. There, the genetic background of the mutant is sufficiently poor that the resulting lineage either has neutral or close-to-neutral fitness. These sites essentially drift neutrally before going extinct; as a result, one should expect extremely rare, nearly private mutations in these populations to be distributed similarly to a population accruing neutral mutations. This separation of regimes is qualitatively different than the classic behavior predicted by sequential selective sweeps, where \textit{any} beneficial mutation is effectively at the high-fitness edge of the population's fitness distribution.

\begin{figure}[h!]
 \begin{center}
 \includegraphics[width=0.8\textwidth]{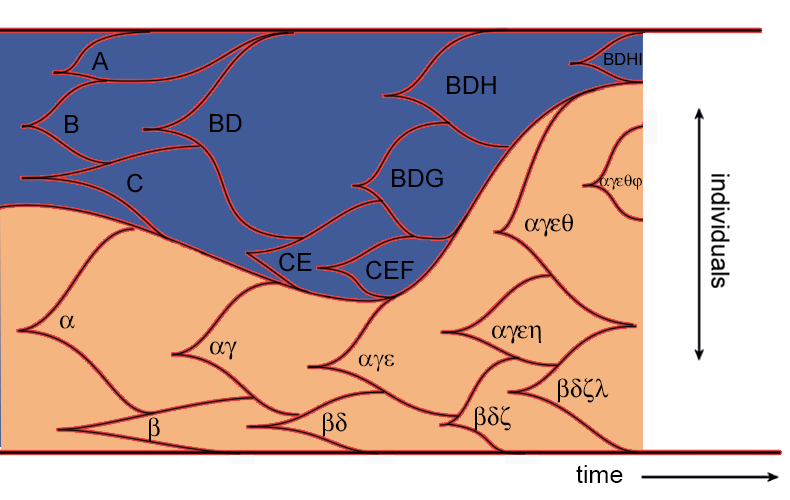}
 \caption{Muller plot demonstrating the underlying evolutionary dynamics of two competing lineages.  Roman capital letters label haplotypes arising on a background that carries a given mutation of interest (blue, darker shading), whereas haplotypes not carrying that mutation are labelled with Greek letters (orange, lighter shading). The frequency of the mutation fluctuates according to the stochastic introduction and expansion of a relatively small number of very fit haplotypes on the two competing backgrounds. \label{mullerplotmore}}
 \end{center}
 \end{figure}

Now, considering only the dynamics of mutants founded on ``good'' genetic backgrounds, we may immediately discern some properties of mutations in these rapidly evolving populations. First, a mutation founded in the high-fitness nose of the distribution will expand in parallel to the growth of its founding fitness class, defined as the subpopulation of mutants with identical (or, for a distribution of fitness effects, close-to-identical) relative fitnesses. Eventually, that founding class comes to dominate the population, at which point the frequency of the mutation in that class is a good approximation to its frequency in the population as a whole. However, in the meantime the lineage of the original mutant has also generated additional beneficial mutations. The future frequency of the mutant will fluctuate depending on how quickly it has done so, relative to its competition (see Figure \ref{mullerplotmore}).

Given enough time, fitter and fitter classes take their turn in expanding and dominating the population, while classes at the low-fitness end of the distribution diminish and go extinct. Eventually, the original founding class of a mutation becomes among the least fit in a population. At this point, the mutation (if it is not already vanished in any existing class) is present among every stratum of fitness in the population, and its fate and future dynamics no longer depend on its selective effect. The original selective benefit of the mutation was only significant in that it pushed the site to an appreciable frequency within its founding fitness class, when that class inhabited the high-fitness nose of the fitness distribution. The dynamics of the mutation after the expansion and domination of its founding class are wholly determined by genetic draft --- i.e., the accrual of further mutants on the genetic background carrying the mutation. This is only dependent on the frequency of a mutation in a given fitness class, and not on whether the actual mutation is beneficial, neutral or deleterious. In what follows, we will flesh out these ideas and others in greater rigor, characterizing the effect of draft in describing trajectories and fates of beneficial and neutral mutations.

\section{Model}

%introduce assumptions - no epistasis or FDS, single selective benefit, strong selection, strong mutation regime so that Ns  \gg 1, s/Ub \gg 1 (subpopulations of various fitnesses)

We study the evolution of a large asexually reproducing population of constant size $N$, using the model introduced in \citet{DF} (summarized below). This model assumes that beneficial mutations of a single fitness effect $s$ occur at a constant beneficial mutation rate $U_b$ per genome and are drawn from an effectively infinite number of possible sites. The use of a single fitness effect allows the fitness of an individual in the population to be described solely by the number of beneficial mutations $k$ it carries, with the absolute fitness of an individual given by $w_k = (1 + s)^k \approx 1 + ks$ for $s \ll 1$. More complicated effects such as frequency dependent selection and epistasis are neglected in this analysis (although the model is easily modified to include some simple sign epistasis, see Discussion).
%Such a modification allows us to extend our analysis to the case of fitness landscapes featuring multiple disjoint evolutionary trajectories.
Finally, we assume that the population is evolving in the strong selection, strong mutation regime. Specifically, this means that $Ns  \gg 1, s/U_b  \gg 1, N U_b  \gg 1$; meaning that the selective forces, selective forces relative to mutations, and incoming mutations per generation are all large.

%introduce model, obligatory picture -> drift, mutation, selection; for more details see DF 07
%Justiification for high-fitness nose and deterministic bulk

In the next few paragraphs we will review the primary features of this model that are pertinent to our analysis, which are justified in detail by \citet{DF}. This model describes the population as a travelling wave in fitness space, wherein the deterministic evolution in the bulk of the wave is combined with a careful stochastic treatment of the birth and fluctuation in lineage sizes of mutants at the high-fitness nose of the distribution. Specifically, the population is characterized according to the number of individuals $n_k$ in each fitness class $k$, where the term \textit{fitness class} refers to the class of individuals carrying $k$ beneficial mutations. At each generation, $n_k$ changes according to the effects of genetic drift, incoming and outgoing mutations, and selection. If $n_k$ is sufficiently large, then selection trumps the effects of the other two evolutionary forces, and the rate of change of $n_k$ is
\[\frac{d n_k(t)}{dt} \approx n_k (ks - \langle ks \rangle ),\]
where $t$ is taken in units of generations and $\langle ks \rangle $ is the (time-dependent) mean fitness of the population.
A fitness class will enter this regime of deterministic growth shortly after the effect of selective forces overcomes the effect of drift, which occurs shortly after the entire fitness class reaches a population size $\sim 1/(k - \langle{k}\rangle)s$, at which point we say that it is \textit{established}. Given our assumption that $s \gg U_b$, the probability for a fitness class that has not yet established to generate a more fit establishing lineage %(i.e, through an additional beneficial mutation in the progeny of one of its individuals that is lucky enough to establish)
 is extremely low. Thus, the population is well described by a deterministically growing/shrinking set of fitness classes $\{ n_{k_{min}}, n_{k_{min}+1}, ...., n_{k_{max}-1} \}$ and one stochastically fluctuating class $\{ n_{k_{max}} \}$, where $k_{min}, k_{max}$ are defined to be the minimum/maximum $k$ s.t. $n_{k_{min}}, n_{k_{max}} \neq 0$ (see Figure \ref{fitnessclasses}A).
%one could consider the evolution of an initially clonal population (transient effects) but we are interested in the case where there is some steady distribution of fitnesses set by equilbrium between the influx and growth of highly fit mutants (increasing the width of the distribution) and the advancement of the mean (decreasing this width).

Although in principle one could consider the transient dynamics by which an initially clonal population attains a steady distribution of relative fitnesses, we are instead interested in the regime where this equilibrium distribution has already been reached and is maintained over timescales long compared to the typical establishment time of a new fitness class. In other words, the population has been evolving long enough to attain some typical steady state fitness profile, but not long enough to begin to deplete the supply of beneficial mutations (which validates our infinite sites approximation). In this case, the width of the distribution is set by an equilibrium between the influx and growth of highly fit mutants (which increases the width of the distribution) and the advancement of the mean fitness (decreasing this width). The size of this width $q$ (defined as the mean number of mutations between the mean fitness class and the largest not-yet-established class) is given to a good approximation by
\[ q \approx \frac{2 \log(N s) }{\log(s/U_b)}, \]
with higher order corrections given in \citet{DF}.
Similarly, $\tau_q$, defined roughly as the random variable denoting the time between the establishment of one fitness class and the next, has expectation value
\[ \langle \tau_q \rangle = \frac{1}{(q-1)s} \log \left[ \frac{s}{U_b} \frac{q \sin(\pi/q)}{\pi e^{\gamma_E/q}} \right], \]
with $\gamma_E \approx 0.577$ the Euler gamma constant.

A more accurate derivation of the true time between establishments, accounting primarily for the non-negligible effect of incoming mutations shortly after a class establishes, is derived in \citet{Brunet1}, whereas a more careful discussion of the correct interpretation of $\tau_q$ is given in \citet{DF}. However, the precise distribution of $\tau_q$ is not important for our analysis, since throughout the rest of this paper we will take time in units of fitness-class establishments.

If $q$ is not small, the mean relative fitness of any individual class does not change too much in the course of one establishment time. In this case, the dynamics of fitness classes are excellently approximated by a \textit{staircase model}, in which the fitness of every class is held constant over the course of the establishment of a new class. When the current most fit class establishes, the fitness of every class is shifted downwards by $s$ and the process repeats. At the time of establishment of each new class $n_i$, the number of individuals in a given class $k \leq i$ is typically
\begin{equation}
n_k \approx \frac{1}{qs} e^{s \langle \tau_q \rangle (q(q+1)- (k- \langle k \rangle)(k - \langle k \rangle+1))/2} \label{3}
\end{equation}
where $\langle k \rangle$ is the mean number of beneficial mutations carried by an individual.
Note that by averaging over all times we recover a Gaussian distribution with variance $\sigma^2 = v = s/\langle \tau_q \rangle$, where $v$ is the rate of adaptation.

\begin{figure}[h!]
  \begin{center}
  \includegraphics[width=0.8\textwidth]{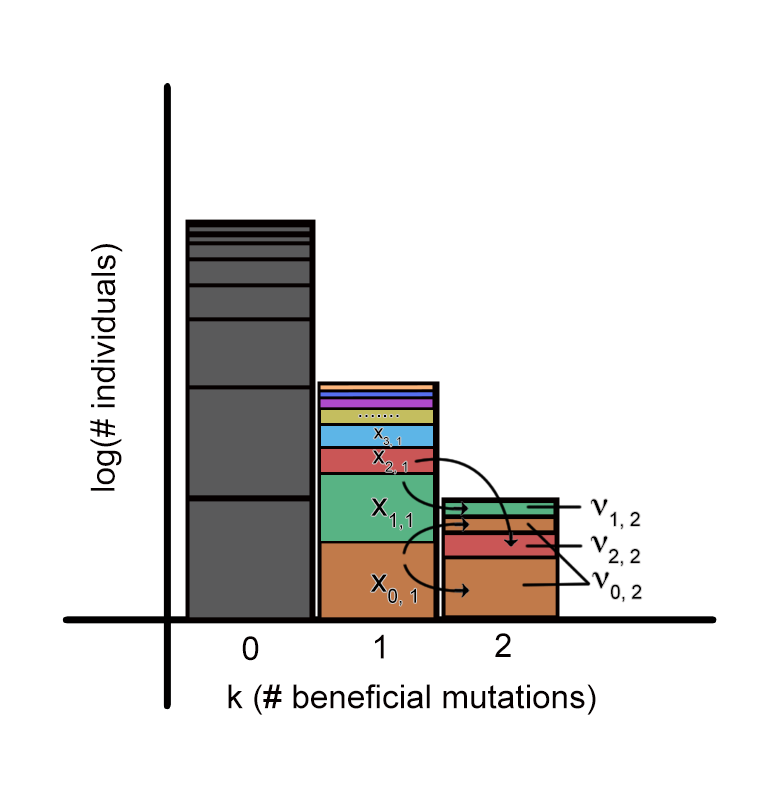}
  \caption{Illustration of the stochastic process by which lineages in a class $k-1$ contribute to a class $k$ through further beneficial mutations. Class $1$ is broken down into haplotypes $\{0, 1, 2,...\ell \}$ at frequencies $\{x_{0,1}, x_{1, 1},..., x_{\ell, 1}\}$ within class $1$, which grow deterministically. At the instant illustrated here, haplotype $0$ (at frequency $x_{0,1}$ within class $1$) has generated two beneficial mutations. These mutations occurred at two different timepoints in the past, grew stochastically, and now cumulatively form a significant fraction of class $2$. Individuals derived from haplotype $0$ now form a fraction $x_{0,2} = \nu_{0,2}/(\nu_{0,2}+\nu_{1,2}+\nu_{2,2})$ of this class. Analogous observations can be made for the other haplotypes. Before class $2$ begins generating mutants destined to establish in class $3$, the frequencies of its haplotypes will be frozen up to small fluctuations, and the process repeats with a new set of frequencies $\{x_{0,2}, x_{1,2}, ..., x_{\ell, 2}\}$. \label{littlefitnessclasses}}
  \end{center}
  \end{figure}

We are primarily interested in the growth of fitness classes when they are still expanding near the high-fitness tip of the wave, which is where mutations destined to reach appreciable frequencies first occur. Thus, we would like to examine the growth of the class $k=k_{max}$, i.e. the largest not-yet-established class. Given such a class, setting $t=0$ at the time of establishment of the \textit{previous} fitness class $k-1$,
%, and assuming that time is sufficiently short so that the fitness of this class does not shift too much,
the number of individuals in the leading class $k$ (or simply the \textit{lead}) at short times can be written as
\begin{equation}
 n_k(t) = \frac{e^{q s ( t- \tau_q)}}{qs} ,
\label{1}
\end{equation}
which can equivalently be recast as
\begin{equation}
 n_k(t) = \frac{\sigma  e^{q s ( t - \langle \tau_q \rangle) - \gamma_E/(q-1)} }{qs} .
\label{2}
\end{equation}
In formulation \eqref{1}, the stochasticity of the growth is encapsulated in the establishment time $\tau_q$; in formulation \eqref{2}, the random variable $\sigma$ encodes how much the class deviates from its typical growth at long times.  The random variable $\sigma$ has the simple generating function (obtained by a transformation of the generating function of $n_k(t)$ given in \citet{DF}),
\begin{equation} \langle e^{- z \sigma} \rangle = e^{-z^{1-1/q}}. \label{gsigma} \end{equation}
\indent Note that $\sigma$ is singularly well suited for extracting the contribution of independent lineages. To see this, we note that $n_k(t)$ denotes the growth of class $k$ given that it is fed by an exponentially (and deterministically) expanding class
\[n_{k-1}(t) = \frac{1}{qs} e^{(q-1)st}\]
which is supplying mutants to class $k$ of fitness $qs$ at a rate $U_b$ per genome per generation. To probe the contribution of particular haplotypes from class $(k-1)$, one could decompose the growth of class $(k-1)$ into
\[ n_{k-1}(t) = \frac{ \sum_{\ell} x_{\ell, k-1}}{qs} e^{(q-1)s t}, \ \ \ \  \sum_{\ell} x_{\ell, k-1} = 1, \] where $x_{\ell, k-1}$ is the fraction of the class $k-1$ constituted by haplotype $\ell$. In this case, the growth of class $k$ may be written as
\[ n_k(t) = \frac{\sigma}{qs} e^{q s ( t - \langle \tau_q \rangle) - \gamma_E/(q-1)} = \frac{\sum_\ell \nu_{\ell, k}}{qs} e^{q s ( t - \langle \tau_q \rangle) - \gamma_E/(q-1)}, \]
where $\nu_{\ell, k}$ is the random variable denoting the contribution from haplotype $\ell$ to class $k$  from class $k-1$. Since it is \textit{not} necessary that $\sigma = \sum_{\ell} \nu_{\ell,k} =1 $, $\nu_{\ell,k}$ is not the frequency of the lineage derived from haplotype $\ell$ in class $k$. Rather, $\nu_{\ell, k}$ is the random variable that encodes how quickly that lineage establishes and expands in class $k$ relative to its typical growth. The frequency of the derived lineage, then, is $x_{\ell,k} = \nu_{\ell,k}/\sigma$. This setup is illustrated in Figure \ref{littlefitnessclasses}. Since each haplotype in class $k-1$ expands exponentially, one could consider $m$ independent feeding processes, where $m$ is the total number of haplotypes in class $(k-1)$. From this, it is straightforward to show that
\begin{equation}
 \langle e^{-z \nu_{\ell, k}} \rangle = e^{-x_{\ell, k-1} z^{1-1/q}} .
 \label{nul}
 \end{equation}
 Note that the assumption of independence between each $\nu_{\ell,k}$ implicitly assumes that the size of the lead $n_k \ll N$. In this case, feedback effects, whereby the growth of each $\nu_{\ell,k}$ affects the advancement of the mean fitness, and thereby the growth of other $\nu_{\ell,k}$, may safely be neglected. When class $k$ leaves the lead and this assumption begins to break down, the frequency of each lineage $\ell$ derived from the corresponding haplotype in class $k-1$ is already frozen in class $k$ (as we will shortly prove) and the lineage is expanding deterministically.

Some objections to this formalism might immediately be raised. First, the number of haplotypes $m$ in class $k-1$ is increasing in time due to incoming mutations from the previous class $k-2$. However, we show in Appendix \ref{transitionprobs} that incoming mutations typically stop contributing significantly to a class shortly after it establishes, and certainly by the time that it itself begins feeding establishing mutants to the next class. Thus, while $m$ may be strictly increasing in time, the combined contribution of these new haplotypes affect the frequencies of all other sites only negligibly. Second, a considerable fraction of these haplotypes could be at frequencies $x_{\ell, k-1}$ such that $x_{\ell, k-1} n_{k-1}(t) \lesssim 1/((q-1)s)$, meaning that these haplotypes cannot be modelled deterministically. This objection becomes important when considering fluctuations in the frequencies of haplotypes that are rare in a given class. On the other hand, if a polymorphic site $x_{\ell, k-1}$ is sufficiently common, its growth by the time the class begins supplying establishing mutants may be modelled deterministically. In this case, the above formalism is well suited to predict the contribution of that lineage $\ell$ to subsequent fitness classes.

We now demonstrate that for sufficiently long times, the random variables $x_{\ell, k}$ are distributed independently of time. That this should be so is seen most clearly by the fact that the growth of an individual lineage in class $k$ mirrors the growth of the class as a whole: the number of individuals  in the lead descended from a lineage $\ell$, $\eta_{\ell, k}$, can be written as
\[ \eta_{\ell, k} (t) = \frac{1}{qs} e^{(q-1) s (t- \tau_{\ell, k})}, \]
where $\tau_{\ell, k}$ is the random variable denoting the establishment time of that lineage in the lead $k$. Its frequency in the class, $x_{\ell,k}(t)$ is then
\begin{equation} x_{\ell,k}(t) = \frac{\eta_{\ell, k}(t)}{n_k(t)} = \frac{ e^{(q-1) s (t- \tau_{\ell, k})}}{ e^{(q-1) s (t- \tau_{q})}} = e^{(q-1) s( \tau_q - \tau_{\ell, k})}. \label{4}
\end{equation}
Since both $\tau_q$ and $\tau_{\ell, k}$ are independent of time for long times \citep{DF}, so is the frequency of lineage $\eta_{\ell, k}$.

Thus, a mutation introduced at some time $t=0$, in founding class $k=0$, after $K$ more fitness classes have established, will contain about
\begin{equation}
 n(K \langle \tau_q \rangle ) \approx \sum_{i=0}^K x_{\ell, i} n_i((K-i) \langle \tau_q \rangle) =  \frac{1}{qs} \left( x_{\ell,K}+\sum_{i=0}^{K-1} x_{\ell,i} e^{(q-1)s \langle \tau_q \rangle} e^{(q-2)s \langle \tau_q \rangle}...e^{(q-(K-i))s \langle \tau_q \rangle}  \right)
\label{trajectory}
\end{equation}
individuals, where $x_{\ell,i}$ are the equilibrium frequencies that the mutation attains in classes $i$, $0 \leq i \leq K$, and $n_i(t)$ is the size of the class $i$ given that it established at $t=0$.

In what follows, since we are only interested in the dynamics of one particular lineage at a time, we will drop the $\ell$ subscript and set the initial fitness class (i.e., the fitness class in which the lineage frequency first begins to be tracked) at $k=0$. The strengths and realm of validity for all of the above assumptions have been studied by ourselves and others in previous work \citep{DF, DWF, FisherMagnumOpus, Brunet1, Rouzine2}.

%Note weaknesses of this approach

\section{Transition Probabilities}

From the considerations of the previous section, we see that evolutionary dynamics in these populations are driven by two factors: the deterministic growth and decay of existing clones --- governing the short time dynamics --- and the stochastic introduction and expansion of new super-fit mutants, which govern the population's long-term evolution.

Once established, a new, high-fitness clone is destined to grow, stagnate and diminish deterministically according to $\eta_k(t) = x_k n_k(t)$. At any moment in time, the population can be divided into many such expanding and contracting ``bubbles'', which fully determine frequency dynamics over short timescales of $\mathcal{O}(\tau_q)$. However, many of the interesting long term dynamics are determined by the stochastic origination and establishment of super-fit mutants from these deterministically expanding clones, which drive the success or failure of particular lineages, mutations, or entire evolutionary trajectories.
%FIGURE 3 INTRODUCED HERE:

\begin{figure}[h!]
  \centering
  \includegraphics[width=12cm]{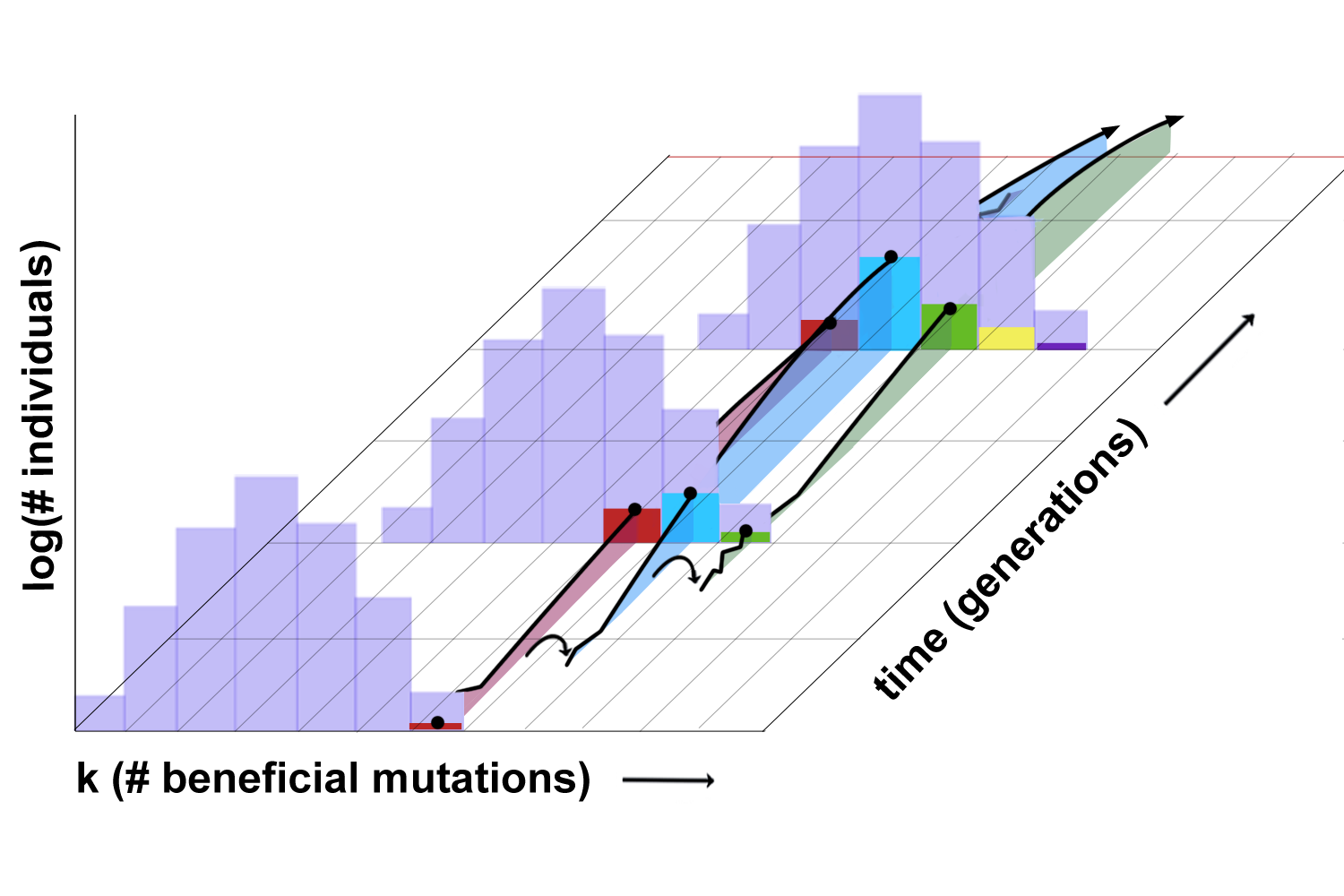}
  \caption{The frequency trajectory of a successful mutation on a high-fitness background. The population's fitness profile is shown for three timepoints, along with one haplotype present in each class. A mutation founded in the first timepoint (labelled in red) expands and constitutes a significant fraction of its fitness class in the second timepoint. As the population adapts, the founding fitness class begins to shrink, along with the number of individuals in the lineage comprising that class. However, this lineage generates a beneficial mutant (shaded in blue) between the first and second timepoints, as shown by the first arrow. This mutant drifts randomly for a short while before establishing and eventually comprising a significant fraction of the next fitness class. In this way, the original successful mutation `jumps' into the next fitness class and is potentially able to avoid extinction. This process repeats, as the second (blue) lineage generates another successful (green) mutant that comes to populate a significant fraction of the next class, and so on.  \label{fig3}}
  \end{figure}

These ideas are expressed more concretely in Figure \ref{fig3}, which shows the distribution of fitness classes at three distinct timepoints. A clone that is about to establish in the first timepoint is growing deterministically in the second timepoint and diminishing in the third timepoint, before finally going extinct as the population evolves to higher and higher fitness. Normally, this would mean that the contribution of the clone's lineage is also extinct; however, the lineage avoids this fate by jumping into the next fitness class through the creation of a new, super-fit mutant when the class is still small and expanding very rapidly. This new mutant establishes and expands, and because it occurs very early, comes to form a significant fraction of the next fitness class, as exemplified in the second timepoint of Figure \ref{fig3}. Although only one of these jumps is shown in Figure \ref{fig3}, a given lineage may jump many times into the next class, with each successive jump, on average, contributing a smaller and smaller fraction of individuals to that class. This new clone will then deterministically expand, contract, and go extinct in the new class, although its lineage may survive by jumping into the next fitness class sufficiently early to eventually constitute a significant fraction of that class, shown by the second jump in Figure \ref{fig3}. The process continues ad infinitum, until the sum of all the contributions of a given lineage to a class vanishes or constitutes the entire class, in which case the lineage is then destined to go extinct or sweep, respectively.
%FIGURE 4 INTRODUCED  HERE:

\begin{SCfigure}
  \includegraphics[width=0.6\textwidth]{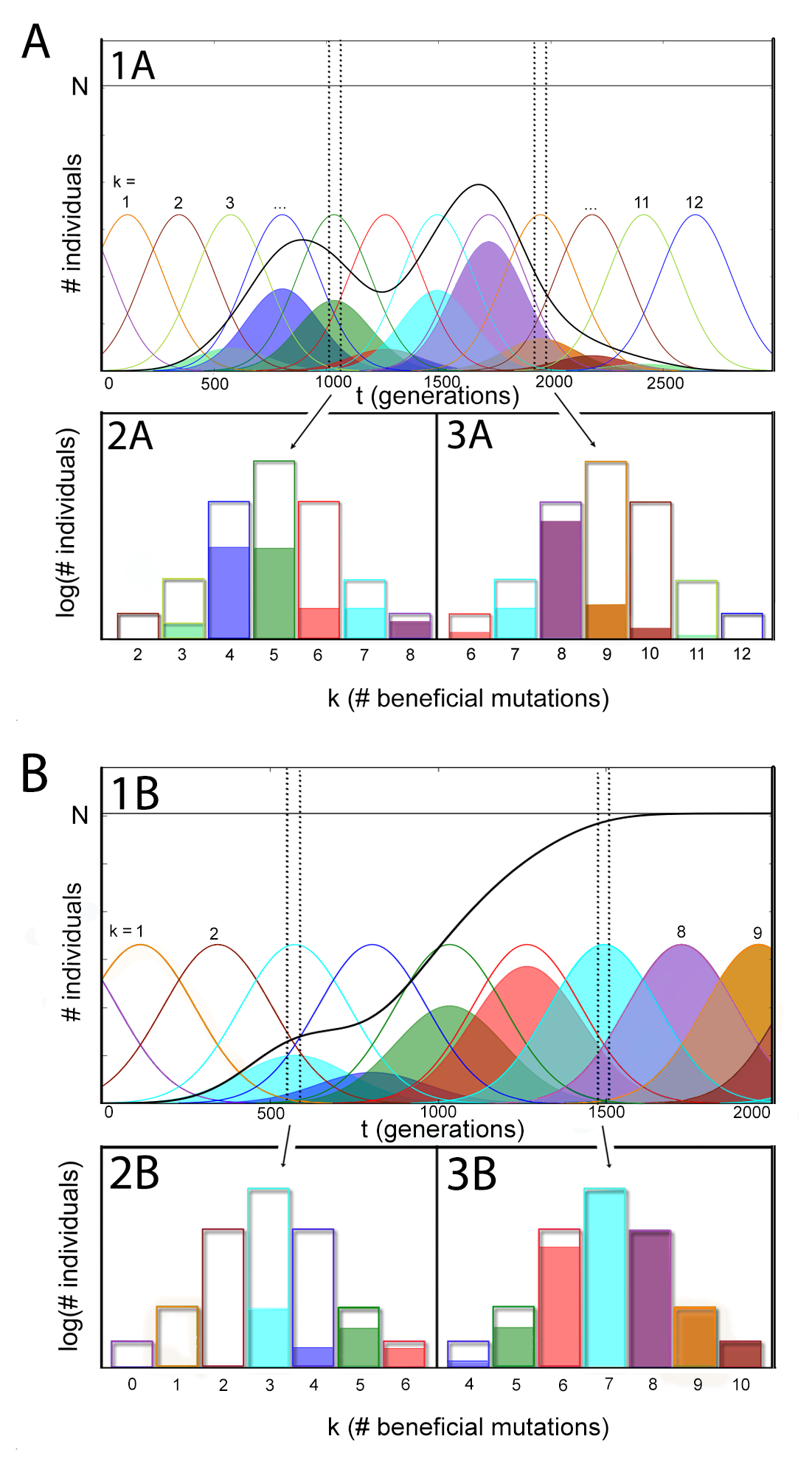}
  %\centering
  %	\subfigure
  %{		\includegraphics[width=9cm]{fig1finished.png}
  %}
  %\hspace{1cm}
  %	\subfigure
  %{		\includegraphics[width=9cm]{fig2finished.png}		
  %}
  \caption{\small The underlying dynamics and observed frequency of a particular mutation in a population with $q=3$, from introduction at $t=0$ to extinction ($A$) or fixation ($B$). The thick black line in $1A, 1B$ denotes the number of individuals carrying a given mutation, and the straight black line at the top of $1A,1B$ represents the total number of individuals in the population. Each solid, colored curve represents a fitness class whose dynamics in time are deterministic (up to fluctuations in the time of establishment of each class, suppressed here for clarity). The shaded region underneath each class denotes those individuals carrying the mutation in that class, with the observed frequency simply the sum of all the contributions from each class. The frequency of the mutant from one class to the next is determined according to the transition probability $\rho$. The profile of the fitness wave at two instants in time is shown in $2A, 3A, 2B, 3B$, with the color of each fitness class matching with $1 A, 1B$. \label{fig4}}
  \end{SCfigure}

The key distribution describing these dynamics is the jump probability $\rho(x_k | x_{k-1},..., x_0)$, the probability of finding a lineage at frequency $x_k$ in fitness class $k$, given that it was at frequencies $x_{k-1}, x_{k-2}, ... x_0$ in the $k$ previous fitness classes. Essentially, $\rho(x_k |  x_{k-1},..., x_0)$ gives the probability distribution of the sum of the frequencies of each clone in fitness class $k$ that originated from a lineage at some frequency $x_0$ in fitness class $0$ and jumped through $k-1$ intermediate classes. This transition process, along with the resulting frequency dynamics of a lineage in the population as a whole, is demonstrated in Figure \ref{fig4} for two independently evolving populations. Under our particular model, the derivation of $\rho(x_k | x_{k-1},..., x_0)$ becomes much simpler because the frequency of a lineage at fitness class $k$ is only determined by its frequency at fitness class $k-1$. This is equivalent to the statement that when a class begins feeding establishing mutants to the next class (i.e., when the jumps in Figure \ref{fig3} occur), the frequencies of lineages in the feeding class are already frozen. In this case, one may consider the frequencies of lineages evolving in analogy to the entire population:  the frequencies of lineages in the classes $\{ n_{k_{min}}, n_{k_{min}+1}, ....n_{k_{max}-1} \}$ are frozen, and the frequencies of mutants in the lead $\{n_{k_{max}} \}$ are fluctuating. Thus the transition process is a Markov chain, meaning that the long-time frequency of a lineage in fitness class $k$, $x_k$, is simply $\rho( x_k | x_{k-1}, ... x_0) = \rho(x_k| x_{k-1})$.  Of course this viewpoint sacrifices precision for the sake of clarity, since frequencies of lineages in the lead will continue to fluctuate after the lead establishes. It is only when a class typically begins to supply \textit{establishing} mutants to the nose that the frequencies of its lineages will be frozen. Regardless, there will typically be one class with fluctuating frequencies (either the lead or next-to-lead class shortly after it establishes) and the rest of the population with lineage frequencies already frozen. This is by no means an obvious assumption, and is discussed in detail in Appendix \ref{transitionprobs}.

In Eq. \eqref{nul} we found that the generating function of $\nu_k$ (the contribution of a particular lineage to a fitness class $k$, given that the frequency of the lineage in class $k-1$ is $x_{k-1}$) is
\[ \langle e^{-\nu_k z} \rangle = e^{-x_{k-1} z^{\alpha}} \]
with $\alpha  = 1 - 1/q$. In class $k$, the lineage will grow as
\[ \eta_k(t)= \frac{x_k}{qs} e^{(q-1) s (t - \tau_q )}  \propto \frac{\nu_k}{qs} e^{(q-1) s (t - \langle \tau_q \rangle )}, \]
where the exact proportionality constant is not relevant for our analysis.

\begin{SCfigure}
  % \begin{center}
   \includegraphics[width=0.6\textwidth]{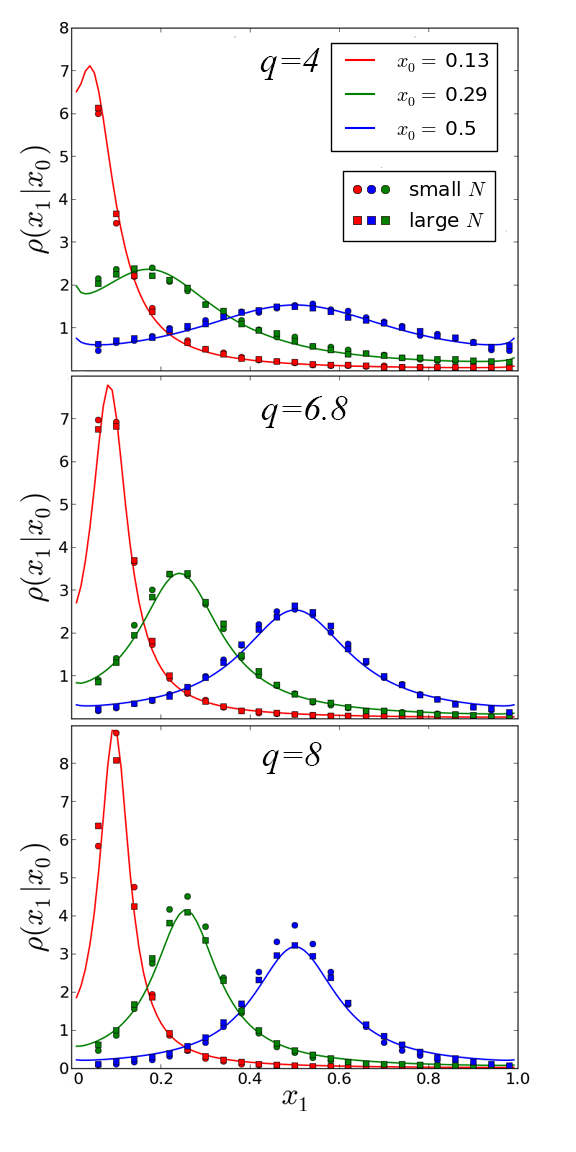}
   \caption{The transition probability $\rho(x_1|x_0)$ plotted for different values of $x_0,$ $q$, and other population parameters $U_b, s,$ and $N$. Solid curves represent the theoretical prediction for the transition probabilities given in  (\ref{rho}). Circles and squares represent the results of forward-time simulations. The circular and square points have identical values of $q$, but differ in population size by a factor of $10$. Circles (small $N$) represent simulations run with parameters $s = 0.1, U_b = 0.001$ and squares (large $N$) are run with parameters $s = 0.01, U_b = 0.0001$. The three plots represent different values of $q$, with $q=4$ derived from the approximate value in the text, and $q=6.8, 8$ derived with higher order corrections given in \citet{DF}. These corrections become more important for higher $q$, and are the reason for the slight discrepancy between theorized and predicted values for small $N$ at $q = 8$. More information about the simulations may be found in Appendix \ref{simulations}. \label{transitionplots}}
   %\end{center}
   \end{SCfigure}

Similarly, we may denote the contribution of that lineage's complement by $\tilde{\nu}_k$ --- that is, the contribution to $k$ from those individuals whose ancestors in class $k-1$ were \textit{not} derived from the chosen lineage. In this case, $\tilde{\nu}_k$ is described by the following generating function:
\[\langle e^{-\tilde{\nu}_k z} \rangle = e^{-(1-x_{k-1})z^{\alpha}} \]
with $\nu_k + \tilde{\nu}_k = \sigma$ and $\sigma$ defined in Eq. \eqref{2}. $\rho(x_k | x_{k-1})$ is then derived from the two generating functions to be
\begin{equation} \rho(x_k | x_{k-1}) = \frac{ \sin(\pi \alpha) x_{k-1}(1- x_{k-1})}{x_k (1-x_k) \pi \left[(1- x_{k-1})^2 \left(\frac{x_k}{1-x_k}\right)^{\alpha} +  x_{k-1}^2 \left(\frac{1-x_k}{x_k}\right)^{\alpha} +  2   x_{k-1} (1-  x_{k-1})\cos(\pi \alpha) \right] }, \label{rho} \end{equation}
which is corroborated by results of forward-time simulations (see Figure \ref{transitionplots}). This is readily extended to $\rho(x_k | x_0)$, the distribution of a given lineage frequency $k$ fitness class steps forward, by the simple replacement $\alpha \rightarrow \alpha^k$. The derivation of (\ref{rho}) and the extension to arbitrary time steps $k$ is given in Appendix \ref{transitionprobs}. Importantly, parameters such as $N, s$ and $U_b$ only factor into this distribution through the parameter $\alpha =  1-1/q \approx 1- \log(s/U_b)/(2 \log(N s))$. Finally, the relationship between the jump probability $\rho$, the measured frequencies of mutations, and the distribution of fitnesses in the population at an instant in time is shown in Figure \ref{fig4}. %This is entirely sensible, since $\rho$ describes the effect of genetic draft on the frequencies of mutations, and it is precisely $q$, the typical number of mutations between the nose and the mean, that quantifies the strength of this evolutionary force in a population.

Although the analytic form of this jump probability is somewhat cryptic, several illuminating properties can be gleaned from its first few moments, derived in Appendix \ref{moments}. Specifically, we find that
%EXPECTATION VALUE
\[ \langle x_k \rangle = x_0, \]
which supports our intuition: the expected value of an allele in future fitness classes is simply its frequency in some reference fitness class. In particular, since $x_{k \rightarrow \infty} \in \{ 0, 1 \}$ (i.e., at long times the allele is either fixed or extinct), this implies that the fixation probability

 \begin{align}
 P_{fix} &=
\int^1_{1-1/N} \rho(x_{\infty} | x_0) dx_{\infty}= 1 \left( \int^1_{1-1/N} \rho(x_{\infty} | x_0) dx_{\infty} \right)+ 0 \left(  \int^{1/N}_{0} \rho(x_{\infty}  | x_0) dx_{\infty} \right) \nonumber \\
&= \langle x_{\infty} \rangle = x_0 \nonumber
\end{align}
Essentially, this means that once a mutation founded at the high-fitness wavefront freezes to a particular frequency in a class, its likelihood of success only depends on its frequency in that class  (at least, without more information about its frequency in fitter classes). %Since a mutation's frequency in a class is correlated to its establishment time relative to other mutations in that class, its timing is of crucial importance with regards to its future success.
We also note that the above formula technically only holds for mutations at high enough frequencies to have necessarily been founded near the distribution's nose class, since mutations founded away from the wavefront have fixation probabilities that are virtually zero. However, since these latecomers never reach more than a negligible frequency in their class, the above formula still describes the fixation probability of a randomly selected mutant reasonably well.

In many cases of biological relevance, the majority of individuals in the population are congregated near the population's mean fitness. As a result, the measured frequency of a mutation tends to be a good estimate of its frequency in the mean class, and thus (barring any information about its frequency in fitter classes) its fixation probability. This fact, along with some of its consequences, will be re-examined in the Discussion.

%Another useful consequence of the first moment is that the measured frequency of a polymorphic site gives the expected value of the frequency of that site in the mean class. This fact is useful in predicting such factors as the sojourn time of mutations.

%VARIANCE

The variance of the jump distribution is
\[ \langle (\Delta x_1)^2 \rangle = \frac{x_0 (1-x_0)}{q}, \]
which is similar in form to the variance associated with genetic drift and single-locus genetic draft. The identities of the first two moments make it tempting to encapsulate the stochastic effects of genetic draft into an effective population size $N_e = q \langle \tau_q \rangle$ (where the factor of $\langle \tau_q \rangle$ arises from rescaling time from fitness class establishments to generations), which might be merged with the additional variance caused by genetic drift. Although intuitively satisfying, this viewpoint tends to be misleading, as the stochastic process described by genetic draft is not diffusive and thus is of a fundamentally different character than drift. Generally, the $n$th moment has the form
\[ \left\langle  x_k^n \right\rangle =  \sum_{i=0}^{n-1} (x_0)^{n-i} \frac{\Gamma(1- \alpha^k+i) \Gamma(n-i)}{(n-1)! \Gamma(1-\alpha^k)} \left( \alpha^{k} \right)^{n-1-i}  \]
where $\Gamma$ denotes the Gamma function, and with higher-order moments falling off as $1/q  \propto 1/N_e$. As a result, the stochastic process is prone to large jumps between frequencies in subsequent classes, even in the limit of infinite population size, $q \rightarrow \infty$. This non-diffusive property of genetic draft has been observed many times for a number of different models \citep{Gillespie1, DWF, Neher1, Neher2}.

\section{Implications for Genetic Diversity}

%genetic draft leaves distinct signatures on the diversity of these populations which are readily measurable, and can be derived using the distribution above

Whereas $\rho$ is not directly measurable by itself --- describing transition probabilities between fitness classes, and not the population as a whole --- genetic draft leaves distinct signatures on the diversity of rapidly adapting populations which are both readily measurable and readily derived from $\rho$. In what follows we derive some implications of the stochastic jump process described by $\rho$ on the site frequency spectra of both beneficial and neutral mutations. Furthermore, it is straightforward to characterize the effects that these jumps have on the structure of genealogical trees (see Appendix \ref{genappendix}), which gives insight into the repeated emergence of Bolthausen-Sznitman statistics in certain aspects of genetic diversity in rapidly adapting populations. We will elaborate on this last point in the Discussion.

\subsection{The site frequency spectrum of beneficial mutations}

One statistic that is strongly affected by the stochastic jumps described above is the site frequency spectrum (SFS) of beneficial mutations $f(x)$, the expected density of mutations between frequencies $x$ and $x+dx$. In rapidly adapting populations, the SFS is partitioned into two regimes: on the one hand, common mutations first arise at the distribution's high fitness nose and are strongly affected by the process of stochastic jumps. On the other hand, nearly private variants, which constitute the majority of beneficial mutations, are overwhelmingly founded near the distribution's bulk and are largely unaffected by this process. Thus, we expect the high and low frequency spectra to be qualitatively different. As a result, the derivation that follows is split into two segments: first, we derive the SFS of common alleles, which are founded at the exponentially expanding wavefront. Since our previous analysis only describes the dynamics of these exponentially expanding nose classes, it does not describe the frequency spectrum of extremely rare, nearly private alleles. Thus, in the second part of this section we make use of a different branching process method to derive the distribution of these plentiful but extremely rare variants.

We begin by deriving the site frequency spectrum of common alleles. For the moment, we may further simplify the problem by considering the site frequency spectrum of mutants in only one fitness class. Since the frequencies of common mutations freeze shortly after a given class establishes, we only need to calculate the distribution of frequencies in a class that is near the nose of the wave. Once the class leaves the nose, the frequencies of these common mutants will be frozen and the class will have the same SFS regardless of its position relative to other classes.

Now, as we have previously argued, almost every common polymorphism was once founded in a class that was at the population's high-fitness nose. Thus, given a fitness class $k$ that is near the distribution's nose, we might decompose the frequency spectrum of mutants in class $k$ according to which class they were founded in. Specifically, we consider mutations founded in class $k$ from class $k-1$ when class $k$ was at the nose; these mutations originate in class $k$. Next, we may consider mutations that originated in class $k-1$ from class $k-2$ when class $k-1$ was near the nose, whose lineages subsequently jumped into class $k$ after acquiring more beneficial mutations; these mutations originate in class $k-1$. Analogously, we can consider the distribution of mutants in class $k$ that originated in classes $k-2, ... k-i$ when these classes were at the distribution's nose. The SFS $f(x)$ in class $k$ is then the sum over all of these distributions.
%For common beneficial mutations, the site frequency spectrum of a given fitness class $k$ may be broken down according to mutations that are $i$ fitness steps old (i.e., according to mutations that were first introduced from the ($k-1$) class to the $k$th class, from the $(k-2)$ class to the $(k-1)$th class, etc., but still persist at intermediate frequencies in class $k$); $f$ is simply the sum over all these distributions.

\begin{figure}[h!]
      \begin{center}
      \includegraphics[width=0.8\textwidth]{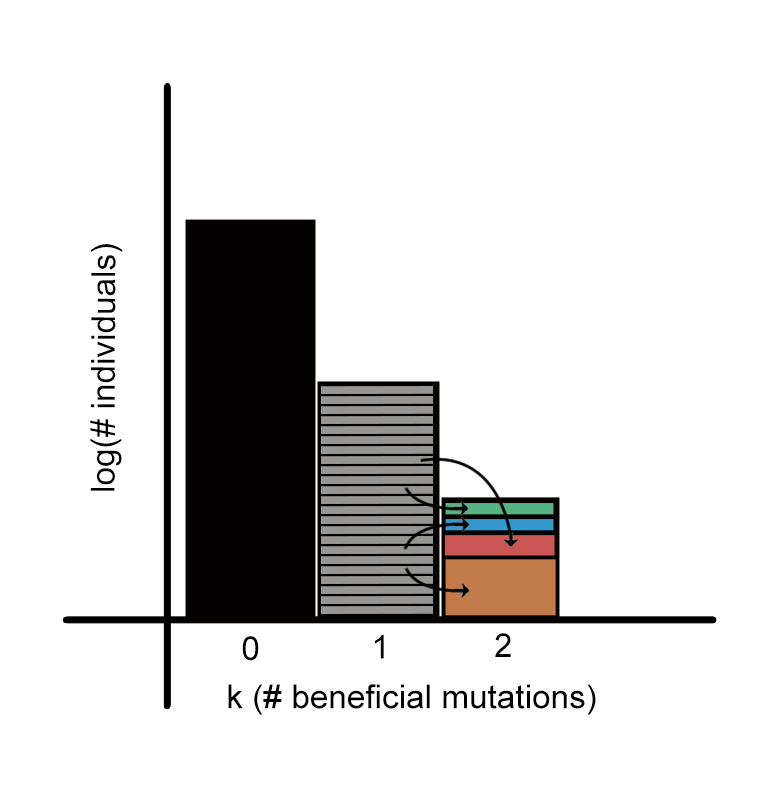}
      \caption{Illustration of the method used to derive site frequency spectra of common mutations and coalescence rates (Appendix \ref{genappendix}). Before it begins supplying establishing mutants to class $2$, class $1$ is decomposed into ($1/\beta$) blocks of frequency $\beta$, such that each $\beta$ is very likely to contribute at most one successful mutant to the next class, but is still large enough to be modelled as growing deterministically. The figure illustrates from which $\beta$-sized block each successful haplotype originated. \label{sfsschematic}}
      \end{center}
      \end{figure}

Consider first the SFS of sites in class $k$ that originate in $k$, $f_1(x)$. We derive the SFS of these ``new'' mutations from the transition probability $\rho$ as follows: first, we observe that class $k-1$ first begins supplying mutants to class $k$ that are \textit{destined to establish} at a time of $\mathcal{O}(\langle \tau_q \rangle)$ since its own establishment. Shortly before this time, one could decompose the growth of class $k -1$ into $1/ \beta$ independently growing blocks of frequency $\beta$ such that $\beta \ll 1/q$. If $\beta$ is sufficiently large, so that $\beta n_{k-1} (\langle \tau_q \rangle) \gg \frac{1}{qs}$, it is valid to model the growth of each block deterministically. In this case, the frequency $x$ in class $k$ of descendants of individuals in block $\beta$ is distributed as $\rho(x | \beta)$. Since $n_{k-1}(\langle \tau_q \rangle) = U_b^{-1}$, we require that $1/q \gg \beta \gg \frac{U_b}{qs}$ for the jumps from each $\beta$-sized block into class $k$ to be well described by $\rho(x| \beta)$. Now, since $\beta \ll 1/q$, it is unlikely that more than two founding mutants originate from the same block. In other words, the contribution of each block to the next class is dominated by the contribution of one most successful mutation, and the probability distribution of the frequency $x$ of this most successful mutation is well approximated by $\rho(x| \beta)$. A diagram demonstrating this setup is given in Figure \ref{sfsschematic}. The expected density of mutations between frequencies $x$ and $x+ dx$ that are introduced from class $k-1$ is then
\begin{equation}
f_1(x) = \frac{\rho( x, \beta) }{\beta} \approx \frac{\sin( \pi \alpha) (1-x)^{\alpha-1}}{\pi  x^{1 + \alpha}} .
\label{rho1}
\end{equation}
Analogously, the distribution of mutations in class $k$ originally arising from the $(k-i)$th class is obtained by the replacement $\alpha \rightarrow \alpha^i$, giving for the total SFS
\begin{align}
 f(x) &= \sum_{i=1}^{\infty} \frac{\sin( \pi \alpha^i) (1-x)^{\alpha^i-1}}{\pi x^{1 + \alpha^i}} \nonumber \\
&\approx \frac{1}{ \pi x (1-x)} \int_0^{\infty}  d\tau \sin( \pi \alpha^\tau) \left( \frac{1-x}{x} \right)^{\alpha^\tau} \nonumber \\
&=  \frac{1}{ \pi x (1-x) \log(q/(q-1))} \int_0^{1}  dz \frac{\sin( \pi z)}{z} \left( \frac{1-x}{x} \right)^{z}. \label{SFS}
\end{align}

%Up to the factor of $\log(q/(q-1))$, this SFS is identical to that associated with the Bolthausen Sznitman coalescent.
Although straightforward to evaluate numerically, this integral has no simple closed form expression. However, a first order Taylor expansion in the sine is a reasonable approximation for $1 - x \ll 1$. This gives
\[ f(x) \approx \frac{1-2x}{\log(q/(q-1)) \log((1-x)/x) x^2 (1-x)}, \ \ \ \ 1-x \ll 1. \]
For $x \rightarrow 1$, \[f \sim ((1-x)\log(1-x))^{-1}.\] Note that this predicts that very high frequency mutations are actually more common than mutations at slightly lower frequencies. This upswing at very high frequency mutations is a widely recognized marker of selection for a number of different models, with and without linkage between sites \citep{Wright2, FayWu, McVeanCharlesworth,  Neher2}. Importantly, it cannot be explained by many commonly studied forms of demographic history, such as population expansions.
 Recently, \citet{Neher2} derived the upswing of the SFS arising from genealogies obeying the Bolthausen-Sznitman coalescent. However, to our knowledge, the excess of extremely common variants arising from  linked beneficial mutations has not previously been derived directly from a specific model of adaptation.

 Another point worth mentioning is the \textit{invariance} of the functional form of this distribution for different parameters $N,s$ and $U_b$. Given that our assumptions about the population hold, the spectra of different populations are \textit{identical} up to a scaling factor that represents the different absolute numbers of common mutations in these populations, which itself is somewhat insensitive to the specific choice of parameters. This observation highlights an inherent limitation in inferring properties of adaptation through the functional form of the site frequency spectrum.

\begin{figure}[h!]
   \begin{center}
   \subfigure{
   \includegraphics[width=0.6\textwidth]{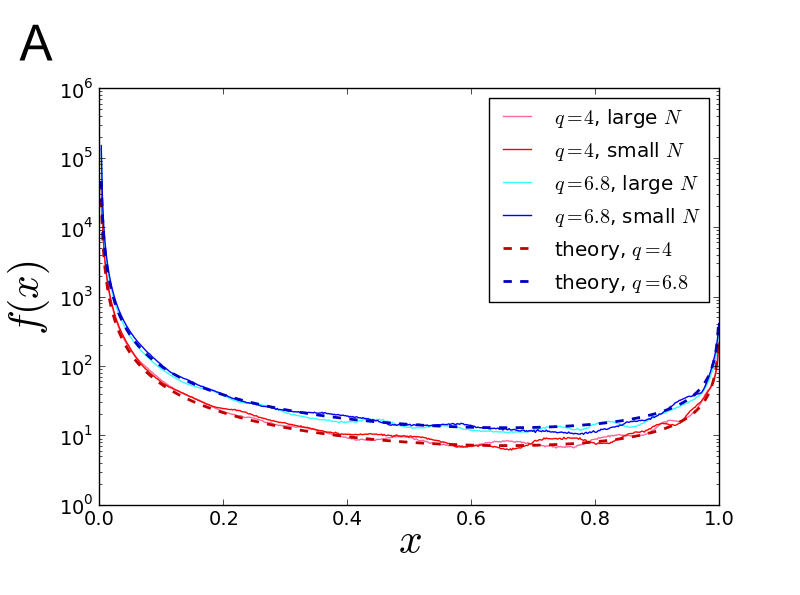}
   }
   \subfigure{
   \includegraphics*[width = 0.6\textwidth]{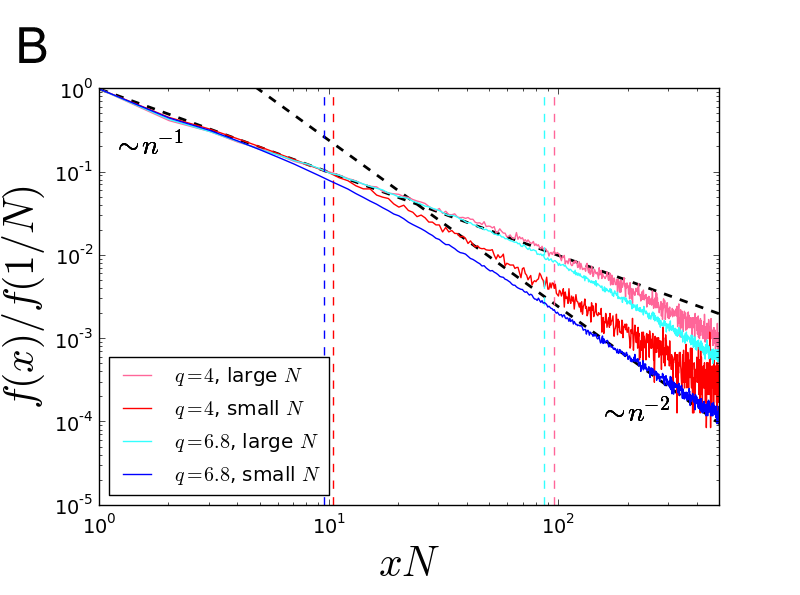}
   }
   \caption{A) The site frequency spectrum of common mutations. The dashed lines are the theoretical predictions from \eqref{SFS} for the two tested values of $q$. Colored lines are frequency spectra derived from the results of 1000 forward-time simulations for each parameter set. For each evolved population, 1000 individuals were sampled from the final timepoint of an initially clonal population evolved for 15000 generations. $s$ and $U_b$ for small and large population sizes are the same as those quoted in Figure \ref{transitionplots}. \label{sfs_all} B) The site frequency spectrum of beneficial semi-private variants (a magnification of (A) in the regime $x \ll 1$). The site frequency spectrum for each parameter set is normalized by the number of purely private mutations (i.e. sites with lineage sizes $n=1$). The  $1/n$ dashed black line is the decay predicted from \eqref{raresfs} for small $n$, and the $1/n^2$ dashed black line indicates the crossover to the (approximately) $1/n^2$ decay predicted for more common alleles. Dashed vertical lines represent the predicted realm of validity of the $1/n$ decay for each parameter set, derived in Appendix \ref{rareSFSappendix}. Colored lines are frequency spectra derived from the results of 1000 forward-time simulations for each parameter set. For each evolved population, the total number of mutations carried by $n$ individuals were explicitly tabulated from the final timepoint of an initially clonal population evolved for 15000 generations. Parameter sets are the same as those quoted in Figure \ref{transitionplots}.  \label{sfs_low_all}}
   \end{center}
   \end{figure}

 We now argue that the above distribution is a good approximation to the population-wide site frequency spectrum, instead of simply the frequency spectrum of mutations in a single class. First, we observe that we can arbitrarily set the distribution $f(x)$ to describe the SFS of the mean class. The above approximation, then, is equivalent to the statement that the SFS of the mean class is a good approximation to the SFS of the entire population.
 %Note that for a long-standing polymorphism (by which we mean a mutation that has been polymorphic for a timescale longer than $2 q \langle \tau_q \rangle$) the above statement is exact: the expected value of a mutation in the mean class is precisely its expected value in the entire population. In fact, its expected value in the population is its expected value in \textit{every} class. The problem, however, arises for recently arising mutations, which may be absent in the mean or less fit classes, but attain substantial frequencies in fitter classes. This would certainly perturb the SFS at lower frequencies.
If $q$ is not too large (the relevant case for many biological populations), then the approximation holds because the vast majority of individuals reside in the mean class at any given time. The contribution of sites from other classes will then be a small perturbation on the SFS of the mean class, particularly relevant at low frequencies (where our approximation breaks down regardless, due to the contribution of mutants not founded at the nose). On the other hand, as $q$ increases, the mean class constitutes a smaller and smaller fraction of the total population. However, the variance of the jumps in the frequencies of mutant sites also decreases in proportion to $1/q$. Thus, while the mean class constitutes a smaller fraction of the total population, sites in adjacent classes tend to shift more slowly than for the case of smaller $q$ (despite the fact that, as we have previously noted, large jumps may still occur occasionally). Thus the approximation should still be valid even in the limit that $q$ is large. The strength of these arguments is corroborated by the close correlation of Eq. \eqref{SFS} with site frequency spectra derived from our forward-time simulations (Figure \ref{sfs_all}).
%%low frequency alleles

As we have previously mentioned, this distribution only describes frequencies of mutations that are founded in the exponentially expanding, high-fitness front of the wave. As such, it fails for rare mutations, which are overwhelmingly dominated by mutations that are introduced when the class is near the mean of the distribution. A different approach is then necessary for understanding the spectrum of these extremely low frequency mutations.

Fortunately, all the difficulties in accounting for effects of genetic draft and the stochasticity of the wavefront are no longer a factor when dealing with these rare variants. As a result, the problem is vastly simplified with the inclusion of two approximations. First, since the lineage sizes of these extremely rare variants are small and destined to go extinct, the lineages can be assumed to experience no further (establishing) beneficial mutations (specifically, this approximation holds if $n y U_b \ll 1$, where $n$ is the lineage size and $y$ the fitness of the mutant created by the lineage). Second, mutations are fed into a given fitness class $k$ deterministically at rate $U_b n_{k-1}(t)$. This holds if $U_b n_{k-1}(t) \gg 1$ (certainly true in the bulk of the distribution), in which case fluctuations of incoming mutants around the expected number are small. Because these lineages never comprise a significant fraction of the population, they can be studied through a standard branching process analysis with a constant death rate $d = 1$ and a time-varying birth rate $b(t) = 1 + y_0 - vt$, where $t$ is time in generations, $y_0$ is the initial fitness of the mutant, and $v$ the mean rate of adaptation of the population. First, we are interested in deriving the expected (time-averaged) number of mutations carried by $n$ individuals with relative fitnesses $y$, $F_{rare}(n/N,y)$. This is obtained by considering the expected number of mutants introduced when a given fitness class was at relative fitness $y_0 > y$, and multiplying by the probability that, in the time it took for the relative fitness of the class to decrease to $y$, the lineage size of any of these mutants has increased to $n$. $F_{rare}(n/N,y)$ is then the integral over all possible landing fitnesses $y_0$.

Clearly, the number of mutants introduced at an initial fitness $y_0$ is simply $U_b N(y_0-s)$ where $N(y)$ is the expected number of individuals at relative fitness $y$. So long as the individual does not reside in the distribution's high-fitness nose, $N(y)$ is well approximated by a Gaussian with variance $v$  \citep{DF}:
\[ N(y) = \frac{N}{\sqrt{2 \pi v}} e^{-y^2/2v}. \] Furthermore, a mutant that is introduced at intial fitness $y_0$ will be at fitness $y$ in a time $t = (y_0-y)/v$. Finally, the distribution in lineage sizes of a mutant with an initial birth rate $1+y_0$ that is decreasing at a rate $v$ per generation is a classic branching process problem that was solved by \citet{Kendall1}. The distribution in lineage sizes is
\[ P_1(n>0,t) =  \frac{ e^{ - y_0 t + \frac{v t^2}{2}}}{(\int_0^t e^{- y_0 \tau + v \tau^2/2} d\tau+1)^2}  \left( 1 - \frac{e^{ - y_0 t + \frac{v t^2}{2}}}{\int_0^t e^{- y_0 \tau + v \tau^2/2} d\tau+1}  \right) ^{n-1}. \]
%The number of individuals with fitnesses $w$ (so long as the individual does not reside in the distribution's high-fitness nose) is well approximated by a Gaussian with variance $v$  \citep{DF}:
%\[ N(y) = \frac{N}{\sqrt{2 \pi v}} e^{-y^2/2v}. \]
%Finally, we define the expected number of sites with lineage sizes $n$ in one fitness class, when that fitness class reaches the relative fitness $y$, to be $F_{rare}(n/N,y)$. This is given by
As a result, $F_{rare}(n/N,y)$ is given by
\begin{equation}
 F_{rare}(n/N, y) = \frac{1}{v} \int_{y}^{(q-2)s}  U_b N(y_0 - s) P_1 (n, (y_0-y)/v) d y_0, \label{frare} \end{equation}
where an arbitrary upper limit of $(q-2)s$ is imposed to restrict to regions where the deterministic supply rate of mutants is guaranteed to hold, with contributions from mutants founded when the class was at fitness greater than $(q-2)s$ already negligible for very rare mutations.
The total SFS of semi-private variants is then obtained by integrating over all final fitnesses $y$: \[F_{rare}(n/N) = \int_{-qs}^{(q-2)s} F_{rare}(n/N, y) dy.\] This integral is examined in detail in Appendix \ref{rareSFSappendix}. Note that in comparing to the Wright-Fisher results we must multiply \eqref{frare} by a factor of $\sim 2$, which reflects the different stochastic dynamics of the branching  process and Wright-Fisher model.

Since the majority of rare variants are observed near the mean fitness of the population, the leading order behavior is well approximated by setting $y=0$. The density of sites, $f_{rare}(x)$ is then approximately
\begin{equation}
  f_{rare}(n/N) \propto F_{rare}(n/N, 0) \approx \frac{U_b N e^{-s^2/(2 v)}}{n \sqrt{2 \pi v}} \propto \frac{1}{n}.  \label{raresfs}
\end{equation}
The conditions for this approximation to hold are that $n \ll \sqrt{2/v} - s/(2v) +1$ and $s/2 \ll \sqrt{2 v}$. Both of these conditions are derived in Appendix \ref{rareSFSappendix}. It is important to observe that these frequencies are at the extremely low end of what is colloquially considered to be a ``rare'' variant, and hence we dub these mutations more precisely as ``nearly private'' or ``semi-private.'' For the frequencies commonly measured in a reasonably sized population sample, rare but non-singleton variants will still decay as $1/x^2$, and the effect of this skew for nearly private mutations will manifest itself as a smaller number of singletons than that predicted simply by the $1/x^2$ extrapolation.

The frequencies of mutant sites thus fall into two regimes. Common alleles founded at the wavefront have distributions similar to those of exponentially expanding populations. Conversely, semi-private variants are largely founded recently in the past and in the bulk of the distribution, and as a result exhibit a neutral SFS. This latter property is only to be expected, since a mutant landing in the mean fitness class has neutral relative fitness by definition, regardless of the specific fitness effects of the mutations it carries. These findings are supported by our forward-time simulations, demonstrated in Figure \ref{sfs_low_all}.
%there is some cross-over region between the two regimes. when does this happen? overlap?

There is necessarily some crossover region between the regime of more common alleles and semi-private variants, occurring in the region of landing fitness $y_0 \sim (q-2)s$. Near these fitnesses, newly founded sites are no longer well described as originating in the expanding, high-fitness front of the wave; however, over the course of their existence a sufficiently significant number of them may reach large enough lineage sizes to be affected by draft. In this case, the distribution of site frequencies in fitness classes near the population's bulk is skewed by mutations that were not founded near the nose, but still jumped into fitter classes. Such lineages will certainly contribute a potentially non-negligible number of sites at the rare end of the spectrum (at lineage sizes larger than the ``semi-private'' ones we have studied). However, because they do not change the results qualitatively, they are neglected in this work. This choice is supported by the rapid crossover between the $1/x$ rare variant decay and the (approximately) $1/x^2$ decay predicted for more common alleles, as exemplified by Figure \ref{sfs_low_all}.

%%% % % % % % % % % % % % % % % % % % % % % % % % % % % % % % % % % % % % %
% % % % % % % % % % % % % NEUTRAL  SITES % % % % % % % % % % % % % % % % %
% % % % % % % % % % % % % % % % % % % % % % % % % % % % % % % % % % % % %

\subsection{The site frequency spectrum of neutral mutations}

Our method is similarly well suited for calculating the SFS of neutral mutations. Once a mutation is present at some frequency in a given class, its frequency in subsequent classes is \textit{purely} determined by draft. Thus, the only difference between the beneficial and neutral cases is in the distribution of mutations first introduced in a given class when that class was at the distribution's nose, $f_1(x)$. The effect of draft on these mutations $i$ classes later is then obtained by convolution with the jump probability,
\[ f_{neut,i>1}(x) = \int_0^1 f_{neut,1}(y) \rho_{i-1}(x|y) dy, \]
where $\rho_{i-1}(x|y)$ denotes the probability density of a mutation at frequency $x$ in a class $i-1$, given that it was at frequency $y$ in class $0$. The total site frequency spectrum is then obtained by summing over all timepoints, corresponding to all possible originating classes:
\[f_{neut}(x) =\sum_{i=1}^{\infty} f_{neut, i}(x). \]

\begin{figure}[h!]
   \begin{center}
   \subfigure{
   \includegraphics[width=0.6\textwidth]{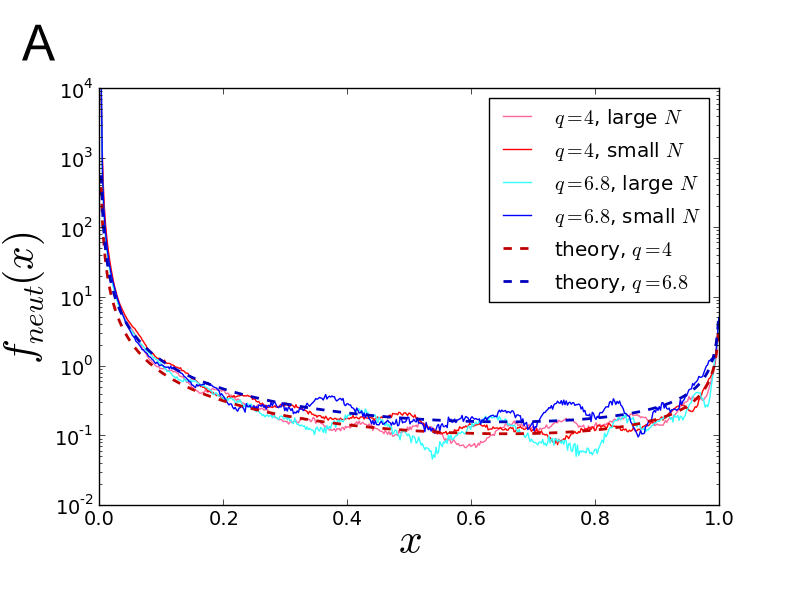}
   }
   \subfigure{
   \includegraphics[width=0.6\textwidth]{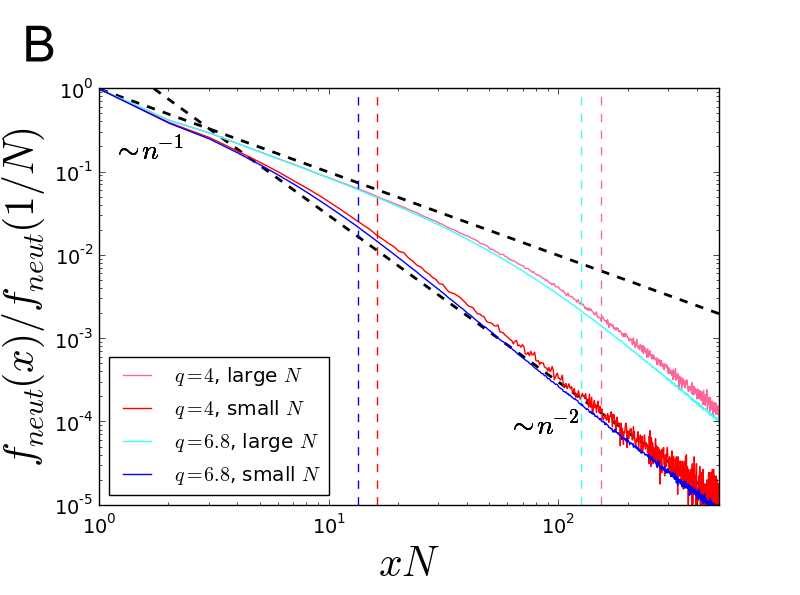}
   }
   \caption{A) The site frequency spectrum of common neutral mutations. The dashed lines are the theoretical predictions using the exact scaling in \eqref{scaling} for the two tested values of $q$. Colored lines are frequency spectra derived from the results of forward-time simulations for each parameter set. For each evolved population, 1000 individuals were sampled from the final timepoint of an initially clonal population evolved for 15000 generations. Parameter sets are the same as those quoted in Figure \ref{transitionplots}, and additionally $U_n = U_b$ for each parameter set. \label{sfs_all_neutral} B) The site frequency spectrum of neutral semi-private variants  (a magnification of (A) in the regime $x \ll 1$). The site frequency spectrum for each parameter set is normalized by the number of purely private neutral mutations (i.e. neutral sites with lineage sizes $n=1$). The  $1/n$ dashed black line is the decay predicted from \eqref{raresfsneutral} for small $n$, and the $1/n^2$ dashed black line indicates the crossover to the (approximately) $1/n^2$ decay predicted for more common neutral alleles. Dashed vertical lines represent the predicted realm of validity of the $1/n$ decay for each parameter set, derived in Appendix \ref{rareSFSappendix}. Colored lines are frequency spectra derived from the results of forward-time simulations for each parameter set. For each evolved population, the total number of mutations carried by $n$ individuals were explicitly tabulated from the final timepoint of an initially clonal population evolved for 15000 generations. Parameter sets are the same as those quoted in Figure \ref{transitionplots}, and additionally $U_n = U_b$ for each parameter set. \label{sfs_low_neutral}}
   \end{center}
   \end{figure}

In Appendix \ref{neutralsfsappendix} we derive that $f_{neut,1}(x) \approx U_n/((q-1)s x^2)$ to leading order. Similarly, $f_{ben,1}(x)$, the distribution of newly introduced beneficial mutations in a given class, was derived in \eqref{rho1} to be
\[f_{ben, 1}(x) dx = \frac{\sin(\pi \alpha)}{\pi} \frac{dx}{(1-x)^{1/q} x^{2-1/q}} \rightarrow \frac{dx}{q x^2} \]
for $q \rightarrow \infty$. In this limit, it is true that
\[\frac{ f_{neut,1}(x) }{f_{ben,1}(x)} = \frac{U_n \pi }{\sin(\pi \alpha) (q-1) s } \approx \frac{U_n}{s}.\]
As a result,
\begin{align}
 f_{neut, i}(x) &= \int f_{neut, 1}(y) \rho_{i-1}(x| y) dy \nonumber \\
 &\approx   \int \frac{U_n \pi }{\sin(\pi \alpha) (q-1) s } f_{ben, 1}(y) \rho_{i-1}(x| y) dy = \frac{U_n \pi }{\sin(\pi \alpha) (q-1) s } f_{ben, i} (x) \nonumber
 \end{align}
and by extension (since this holds for each $i$),
\begin{equation} f_{neut}(x) \approx \frac{U_n \pi }{\sin(\pi \alpha) (q-1) s } f_{ben}(x) \rightarrow  \frac{U_n}{s} f_{ben}(x) \label{scaling} \end{equation}
in the limit $q \rightarrow \infty$. Thus, in the limit of rapid adaptation, the neutral and beneficial site frequency spectra differ only by a scaling factor set by the rate of neutral mutation relative to the strength of selection. This relationship demonstrates the fact that, as beneficial mutations become more common, the relative importance of a mutation's intrinsic fitness effect diminishes relative to the quality of its genetic background. Thus, the relative site frequency spectra are roughly set by the rate at which neutral mutations accrue in the nose classes relative to beneficial mutations. Although only strictly true in the infinite adaptation limit, our simulations demonstrate that this approximation is already strong for $q$ as small as $4$ (Figure \ref{sfs_all_neutral}).

To compute the site frequency spectrum of neutral semi-private variants, the derivation follows identically as for beneficial mutations, with $s \rightarrow 0, U_b \rightarrow U_n$. The result may then immediately be written down to be
\begin{equation} f_{neut,rare}(n/N) \propto F_{neut, rare}(n/N, 0) \approx \frac{U_n N}{n \sqrt{2 \pi v}} \propto \frac{1}{n} \label{raresfsneutral} \end{equation}
which holds so long as $n \ll \sqrt{2/v}+1$. This $1/n$ dependence is demonstrated in Figure \ref{sfs_low_neutral}.

\section{Sojourn Times}

So far, we have described the effect of genetic draft on the frequencies of lineages as they jump through fitter and fitter fitness classes, and characterized the effects of these jumps in skewing the resulting beneficial and neutral SFS. Now, we are ready to make predictions regarding fates and trajectories of observed polymorphic sites in these populations.
One of the most important predictions to be made is the time to fixation of any one beneficial allele when compared to the strong selection, weak mutation regime.
%rate is \tau_q for both; \tau_q itself decreases for SSSM regime, but a beneficial allele jumps through many fitness classes before fixing in any one. There are two pertinent questions: how long does it typically take any lineage in a given fitness class to sweep (equivalently, a mutation of some effect $ks$)? Given a mutation at some measured frequency, how long before it sweeps/goes extinct- i.e., its sojourn time?

 %Because every fitness class carries a mutation founded in that class which is eventually destined to fix, mutations fix on average at a rate of $1/ \langle \tau_q \rangle$ per generation. Thus, in both the strong selection, strong mutation (SSSM) and strong selection, weak mutation (SSWM) regimes, the substitution rate of beneficial mutations is simply $1/ \langle \tau_q \rangle$.
 In the strong selection, weak mutation (SSWM) regime, an establishing beneficial mutation usually fixes in its founding fitness class, because in this regime the sweep time of a beneficial mutant is much smaller than the rate at which new beneficial mutations establish. On the other hand, in the strong selection, strong mutation (SSSM) regime, the lineage carrying a particular mutation usually jumps through many fitness classes before fixing in any one. Furthermore, it takes time for the class in which the mutation first fixed to traverse the length of the wave, adding roughly $2 q \langle \tau_q \rangle$ generations before the mutation is fixed in the population.

Thus, in studying the fates of mutations, there are two pertinent questions. First, given a mutation at some measured frequency, how long does it take before the mutation sweeps or goes extinct? Second, given a newly established fitness class, how long does it take \textit{any} mutation introduced in this class to sweep (equivalently, what is the expected time to fixation of a new mutation that is destined to fix)?

%A mutation at very low frequencies will generally drift for a few fitness classes before going extinct. Our method takes into account effects of draft, but not the drifting at very low frequencies. Since Pfix = f, we can calculate the sojourn time up to some small epsilon, beyond which extinction/fixation is nearly certain. The method should start to fail for when epsilon is not above the `establishment threshold' for when a class begins feeding establishing mutants, since at this point the growth of frequency is not deterministic.

Because our method assumes that lineages in the feeding class are frozen once they begin to feed establishing mutants into the next class, our method is poorly equipped to deal with lineages at very high or low frequencies that are strongly affected by drift. Furthermore, the pathologies of our distribution (which, treating $n_k$ as a continuous variable, allows for fractional numbers of individuals) introduces errors in the regime of $n_k \sim \mathcal{O}(1)$. Nevertheless, we can calculate the sojourn time of these mutants by predicting when the frequency of a given mutation is expected to fall above or below a small threshold frequency $\epsilon$. All the above problems may be circumvented if $\epsilon$ is taken to be small, but large enough for the lineage to be established in the feeding class when it begins supplying establishing mutants (roughly, this is fulfilled when $1 \gg \epsilon \gg \frac{U_b}{qs}$). Because $P_{fix} = x$, when a lineage falls below frequency $\epsilon$ or rises above frequency $1-\epsilon$ in a given class, its probability of extinction or fixation is then $1 - \epsilon$, which is nearly certain if $\epsilon$ is sufficiently small. We note that this scenario more accurately imitates what one could measure in an experimental setting, with a sample that is much smaller than the total population size or (if performing whole population sequencing) some finite sequencing read depth. In these experimental scenarios, the absence of a particular polymorphism in such a measurement may not mean that the polymorphism is extinct entirely, but rather that it is unlikely to be present in the population above a certain frequency.
%Sojourn time for any epsilon, given $x_0$, is calculated by X

\begin{SCfigure}
  % \begin{center}
   \includegraphics[width=0.6\textwidth]{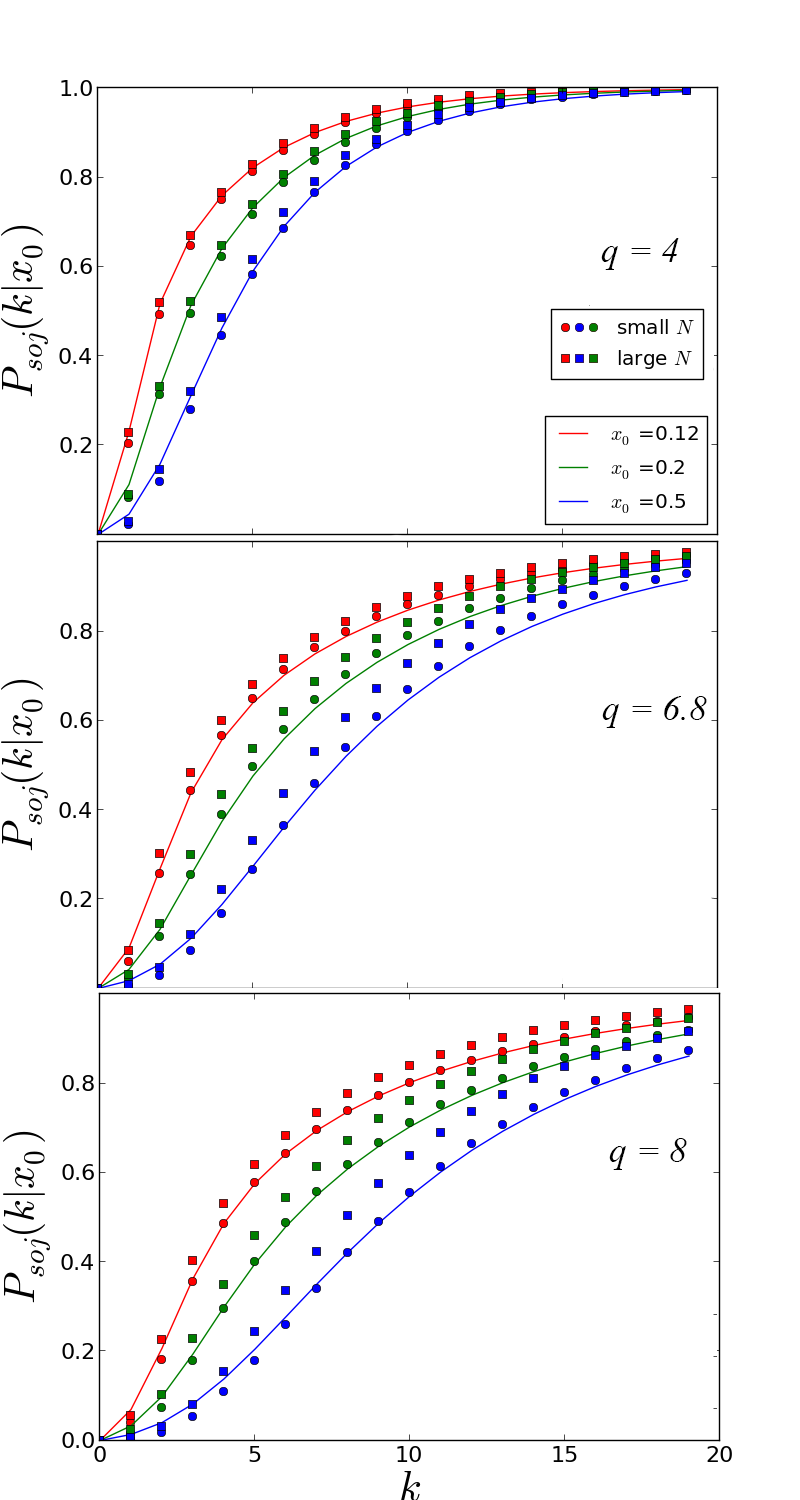}
   \caption{The sojourn probability $P_{soj}(k|x_0)$ plotted for different values of $x_0,$ $q$, and other population parameters $U_b, s,$ and $N$. Solid curves represent the theoretical prediction for the sojourn times given in  \eqref{sojourntimes}. Circles and squares represent the results of forward-time simulations. The circular and square points have identical values of $q$, but differ in population size by a factor of $10$. Circles represent simulations run with parameters $s = 0.1, U_b = 0.001$ and squares are run with parameters $s = 0.01, U_b = 0.0001$. The three plots represent different values of $q$, with $q=4$ derived from the approximate value in the text, and $q=6.8, 8$ derived with higher order corrections given in \citet{DF}. For both simulated and theoretical curves, mutations were considered extinct or fixed in a given class when they fell below or rose above a cutoff of $\epsilon = 0.03$ or $(1-\epsilon) = 0.97$, respectively. More information about the simulations may be found in Appendix \ref{simulations}. \label{sojournplots}}
   %\end{center}
   \end{SCfigure}

The distribution governing the sojourn time $P_{soj}(k | x_0)$ --- the probability that a site at a frequency $x_0$ in fitness class $0$  has a frequency $x_k$ in fitness class $k$ that freezes below the threshold $\epsilon$ (or above $1 - \epsilon$) --- is calculated as
\begin{align}
P_{soj}(k | x_0) &= \int_0^{\epsilon} \rho(x_k| x_0) d x_k + \int_{1-\epsilon}^{1} \rho(x_k| x_0) d x_k  = \int_0^{\zeta} \Pi(\tilde{x}_k | x_0) d \tilde{x}_k + \int_{1/\zeta}^{\infty} \Pi(\tilde{x}_k | x_0) d \tilde{x}_k  \nonumber \\
&=\frac{x_0}{\pi} Re \left[  \int_0^{\zeta}\frac{1}{-i(x_0 \tilde{x}_k + (1-x_0) \tilde{x}_k^{1- \alpha^k} e^{i \pi \alpha^k})} d \tilde{x}_k + \int_{1/\zeta}^{\infty} \frac{1}{-i(x_0 \tilde{x}_k + (1-x_0) \tilde{x}_k^{1- \alpha^k} e^{i \pi \alpha^k})} d \tilde{x}_k \right] \nonumber \\
&= \frac{1}{\alpha^k \pi} \left[ \arctan\left( \frac{x_0 \zeta^{\alpha^k} \sin(\pi \alpha^k)}{1-x_0 + x_0 \zeta^{\alpha^k} \cos(\pi \alpha^k)}\right) + \arctan\left( \frac{(1-x_0)\zeta^{\alpha^k} \sin(\pi \alpha^k)}{x_0 + (1-x_0)\zeta^{\alpha^k} \cos(\pi \alpha^k)} \right) \right] \label{sojourntimes}
\end{align}
where $\rho, \Pi$ are given in \eqref{rho} and \eqref{Pi}, respectively, and $\zeta =\epsilon/(1- \epsilon)$. % Note that this is the probability that a mutation at some frequency $x_0$ in some initial reference class $k=0$ is either (nearly) fixed or extinct $k$ fitness classes forward.
If the initial reference class is assumed to be the mean class, this is the probability that the mutation will be fixed or extinct in the mean class $k \langle \tau_q \rangle$ generations later. It will then require $q \langle \tau_q \rangle$ more generations before the mutation is at a population-wide negligible frequency, corresponding to the time for the mean fitness class to go extinct. Given a mutation at a measured frequency $x_0$ at $t = 0$, the true, population-wide sojourn time in generations is then $\mathcal{P}_{soj}( (k + q) \langle \tau_q \rangle | x_0) = P_{soj}(k | x_0)$.
Both simulated and predicted sojourn times for a number of different parameters are shown in Figure \ref{sojournplots}.

A similar method is used to calculate the expected time for fixation of \textit{any} lineage in a given reference fitness class, which we dub $P_{fix}(k)$. Specifically, $P_{fix}(k)$ is the probability that one of the mutations founded in class $0$ is past the threshold $1-\epsilon$, $k$ fitness classes forward. Using the distribution of \textit{new} mutations given in \eqref{rho1}, the probability that one is at frequency greater than $(1-\epsilon)$ in class $k$ is
\[ P_{fix}(k) = \frac{\sin( \pi \alpha^k)}{\pi} \int_{1-\epsilon}^{1}  \frac{(1-x_k)^{\alpha^k - 1}}{x_k^{\alpha^k + 1}} dx_k \approx \epsilon^{\alpha^k} \]
for $\epsilon \ll 1$ and $k$ large, with the expected fixation time,
\[ \langle k_{fix} \rangle = \sum_{k=1}^{\infty} k \frac{d\epsilon^{\alpha^k}}{dk} \approx q ( \gamma_E + \log(\log(1/\epsilon))) \sim \mathcal{O}(2q) \]
for $q$ large, where the last relation holds for most reasonable values of $\epsilon$. Thus, it takes a time of about $2 q \langle \tau_q \rangle$ for a mutation destined to fix to sweep in any class, and an additional time of about $2 q \langle \tau_q \rangle$ for that class to sweep through the population. Thus the expected sweep time for successful mutations is $\mathcal{O}(4 q \langle \tau_q \rangle) \sim \mathcal{O}(4 \log(s/U_b)/s)$ for $q$ large. Surprisingly, this predicts that as $q$ increases, the time to fix any one mutation (or, more precisely, the time for a mutation to pass some threshold frequency $1-\epsilon$) becomes independent of population size. This condition arises from the fact that, while the time to establish each new class decreases as population size increases (thus accelerating fixation by accelerating the rate of adaptation), this is counteracted by an increase in the number of total fitness classes $q$, increasing both the time for a mutation to fix in any class (by decreasing the variance of $\rho$) and to traverse the bulk of the wave to dominate the population.

\section{Discussion}

We have used a simple, infinite-sites model of adaptation featuring a single beneficial selection coefficient to carefully account for the effects of genetic drift, mutation, selection, and by extension, genetic draft in determining the evolutionary dynamics of polymorphic sites. On short timescales of $\mathcal{O}(\tau_q)$, dynamics of alleles are dominated by the deterministic growth and decay of clonal bubbles according to the relative fitness of each clone. The interesting long-term behavior, however, consists of the creation and establishment of new haplotypes, which drive the fluctuations of polymorphisms over longer timescales. Fortunately, if the population is adapting sufficiently rapidly, these long-term dynamics are dominated by the fates and identities of a small subset of fittest clones. The problem then simplifies to understanding the changing composition of each new, fittest class.
%%%%%%%%%%%%%%%%% Connection to 'classic' draft %%%%%%%%%%%%%%%%%%%%%%

There is a natural connection between our model and the classic model of genetic draft between a neutral locus and a strongly selected locus, first studied by \citet{Gillespie1, Gillespie2}. In these seminal works, the diffusive random walk of a neutral allele is coupled with the stochastic process of a hitchhiking event occurring at rate $R$, which drives the neutral allele to either fixation or extinction. In Gillespie's model, the time required to fix the strongly selected allele relative to the time between substitutions is small enough so that fixation/extinction is assumed to occur instantaneously. One basic result derived under the assumptions of such a model are the first two moments of the stochastic jump process: $\langle \Delta x \rangle = 0, \langle \Delta x^2 \rangle = R x_0 (1-x_0)$, for $\Delta x = x_1 - x_0$ (here, $x_1$ denotes the frequency of the neutral allele after some time of $\mathcal{O}(\langle \tau_{fix} \rangle)$ for the strongly selected locus). In this work, we find that the fundamental properties characterizing draft are preserved in populations where adaptation proceeds with the simultaneous substitution of many sites, instead of the two originally studied. This is particularly evident through the identities of the first two moments of our jump distribution $\rho(x_1 | x_0)$: $\langle \Delta x \rangle = 0$ and $\langle \Delta x^2 \rangle = x_0 (1-x_0)/q$. The first two moments not only have the correct functional dependence on $x_0$, but reduce to previously derived values in the weak mutation limit. This can be seen by considering the fact that our jump probability takes time in units of fitness class establishments, which is equal to the substitution time of beneficial mutations. In the weak mutation limit, we have $q \rightarrow 1$, so that  $\langle \Delta x^2 \rangle \rightarrow x_0 (1-x_0) = R x_0 (1-x_0)$, where $R$ is the substitution rate of beneficial mutations. The replacement $1/q \rightarrow R$ applies to higher order moments as well. Thus our results may be taken to generalize draft to the situation where the time between beneficial mutations is no longer large.
%%%%%%%%%%%%%%%%% Frequency Dynamics in Experiments %%%%%%%%%%%%%%%%%%

\subsection{Quasi-Neutrality over Long Timescales}

When many beneficial mutations segregate simultaneously, frequency dynamics of mutations begin to exhibit a qualitatively different behavior than those in the strong selection, weak mutation regime. After the introduction and establishment of a mutation in the fittest class (which, as we have already argued, is the founding location of the vast majority of successful mutations), a mutation will typically be present in the mean fitness class a time $q \langle \tau_q \rangle$ later. This corresponds to the timescale for the fittest class to become the mean class. Since the mean class consists of the majority of individuals in the population, the measured frequency of the mutant at this time is a good approximation to its frequency in the mean class. Since the frequency of the mutation in subsequent classes is solely determined by genetic draft, its measured frequency at this time is roughly its fixation probability.\footnote{A better approximation is straightforwardly obtained by taking into account the fact that the mutation's frequency in classes below the mean is identically $0$.} Furthermore, at a time of $\mathcal{O}(2q \langle \tau_q \rangle)$, the founding fitness class is the least fit in the population. The mutant's frequency in all classes is then solely determined by draft, whose effect does not depend on the mutation's intrinsic fitness effect. At this point, its measured frequency approaches the expected value of its frequency in the mean class, and thus its fixation probability.

These considerations have important consequences for predicting the fate of polymorphic sites in experimentally evolving populations. In a population that is adapting rapidly enough for a stratification of fitnesses to always be present, sites that are polymorphic for sufficiently long have dynamics that are indistinguishable from neutral mutations at comparable frequencies in these populations, because they are \textit{only} determined by genetic draft. Once the frequency of a mutation is frozen in a particular fitness class, its fitness effect alone is no longer important in determining its future success. This is because it is only the net fitness of the mutant \textit{and} its genetic background that is important in determining its dynamics, and not the fitness effect of the mutation itself. Because it takes a time of $\mathcal{O}(2  \langle \tau_q \rangle )$ for a mutation founded at the nose to occupy all strata of fitnesses in the population, common mutations in these rapidly adapting populations can be thought to have a \textit{relaxation time} of  $2 q \langle \tau_q \rangle$, beyond which their mutational effect decouples from their dynamics.

In short, the significance of this finding is simple: if a mutation --- regardless of its fitness effect --- has been measurable in the population for longer than $2 q \langle \tau_q \rangle $ generations, its fixation probability is equal to its frequency in the population. This approximation should already be quite good (albeit consistently too low due to the inclusion of less-fit classes where the mutation is absent) at a time $q  \langle \tau_q \rangle$.
%effect of draft is stochastic amplification of fluctuations

These considerations provide a parallel to Haldane's formula for the fixation probability of a beneficial mutation in the successive sweep regime, $P_{fix} =  s$. Whereas this formula gives the likelihood of a mutation to escape genetic drift, the probability of a long-standing polymorphism to escape draft is simply $P_{fix} = x$, where $x$  is the frequency of the mutation in the population. The two forces are similar in that they stochastically amplify the fluctuations in the trajectories of polymorphic sites, but the time- and length-scale of the fluctuations caused by draft are much greater.
%%%%%%%%%%%%%%%%%%% Genetic Diversity and the BSC %%%%%%%%%%%%%%%%%%%

\subsection{Relation to the Bolthausen-Sznitman Coalescent}

Having established the effect of genetic draft on the frequencies of polymorphic sites, we then explored the effect that the stochastic jumps characterizing the process have on genetic diversity. In particular, we derived the site frequency spectrum of both beneficial and neutral mutations. As expected, we found that nearly private variants --- which overwhelmingly occur and drift near the mean class --- decay according to the well-known $1/x$ behavior predicted for unlinked neutral sites, whereas common mutations exhibit site frequency spectra more closely related to those of exponentially expanding clones, with the additional feature of an upswing at high frequency that is characteristic of adaptation for many models \citep{Neher2, MesserPetrov}.

Of particular interest is the diversity of populations in the limiting case of $q \rightarrow \infty$, whose genealogies obey the Bolthausen-Sznitman coalescent. Note that in this case, the Bolthausen-Sznitman coalescence rates apply for individuals in the lead, with time taken in units of fitness class establishments (a setup which is elaborated upon in Appendix \ref{genappendix}). Although this was observed previously in \citet{DWF}, in the context of these stochastic jumps the origin of this structure is intuitive: in the high-$q$ limit, the contribution of one individual at the wavefront to the next class falls off as $1/x^2$, which is precisely the form of the offspring distribution that gives rise to Bolthausen-Sznitman statistics. The $1/x^2$ decay is also well-known to describe the frequency spectrum of exponentially expanding populations, and in the case of adapting populations results from the exponential expansion of new mutants at the front of the wave. The Bolthausen-Sznitman coalescent has been associated with a growing number of different adaptive models \citep{Brunet2, Brunet3, Neher2, DWF}, suggesting that such genealogies may be a universal limiting feature of rapid adaptation.

\subsection{Implications for the McDonald-Kreitman Test}

Another useful application of our findings is the ability to analytically correct for the effect of genetic draft on the results of tests for signals of adaptation, such as the McDonald-Kreitman test \citep{McDonaldKreitman}. This test, along with many other widely-used tests for selection, assumes that beneficial mutations are rare and segregate independently. However, both assumptions are invalid for rapidly adapting populations, and new analytical predictions are needed. Fortunately, we will demonstrate that our method provides a simple way to correct for the effect of linkage between beneficial mutations. For the case of the McDonald-Kreitman test, this correction is straightforwardly obtained by accounting for the extra heterozygosity contributed by many simultaneously segregating beneficial mutations.

The McDonald-Kreitman test approximates the fraction $\alpha_{MK}$ of nucleotide substitutions that are adaptive by considering relative quantities of fixed and polymorphic, synonymous and nonsynonymous sites between two diverged populations. The fraction of adaptive substitutions is simply
\begin{equation} \alpha_{MK} = \frac{d_+}{d_n} \approx 1 - \frac{d_s}{d_n} \frac{p_n}{p_s}, \label{MKtest} \end{equation}
where $d_+, d_s,$ and $d_n$ are the adaptive, synonymous and nonsynonymous substitution rates, and $p_n, p_s$ are the numbers of nonsynonymous and synonymous polymorphisms in one of the sampled populations, respectively. The last approximation arises from the assumption that
\begin{equation}
d_+ = d_n-\bar{d} \approx d_n - d_s (\bar{p}/p_s) \approx d_n- d_s (p_n/p_s), \label{d+}
\end{equation} where $\bar{d}$ is the substitution rate of nonadaptive, nonsynonymous mutations and $\bar{p}$ is the number of nonadaptive, nonsynonymous polymorphic sites in the sample. Implicit in \eqref{d+} are several assumptions: first, the rate of nonadaptive, nonsynonymous substitutions in the sample is equal to the rate of synonymous substitutions, scaled by the relative frequencies of nonadaptive, nonsynonymous to synonymous polymorphisms.  This implicitly assumes that deleterious mutations do not fix, that deleterious mutations do not significantly contribute to the measured numbers of nonsynonymous polymorphisms in the sample, and that the population has not undergone any demographic change to skew the distributions of polymorphic sites. Second, it is assumed that the number of nonadaptive, nonsynonymous polymorphisms is precisely equal to the measured numbers of nonsynonymous polymorphisms. In other words, beneficial mutations are rare and fix quickly upon arising; thus, they are rarely present in the population as polymorphisms.

Of course, these assumptions break down when deleterious or beneficial mutations significantly contribute to the number of polymorphic sites and when linkage between sites skews the relative frequencies of mutations. These issues usually result in a measured $\alpha_{MK}$ that severely underestimates the true fraction of adaptive substitutions. A large body of work has been put forward in an effort to correct for this skew \citep{MesserPetrov, Andolfatto1, Charlesworth1, EyreWalker1}. For example, introducing a low-frequency cut-off for measured polymorphisms significantly improves estimates of $\alpha_{MK}$ for the case of many weakly deleterious mutations  \citep{Charlesworth1, Fay1}, since in the absence of genetic linkage few deleterious mutations will ever reach high frequencies. However, few studies have carefully analysed the effect of linkage on $\alpha_{MK}$, and particularly on the effect of linked beneficial mutations. One notable exception is the work of \citet{MesserPetrov}, who used a sophisticated extension of the McDonald-Kreitman test accounting for demographic history and distributions of fitness effects to infer $\alpha_{MK}$ from simulated rapidly adapting populations. The authors uncovered that the inference of a massive population expansion (derived from the site frequency spectrum of the sample) resulted in superior estimates of $\alpha_{MK}$, although no such expansion ever occurred in the simulation. The intuition guiding this finding is that the site frequency spectrum of a population undergoing rapid adaptation resembles that of a population undergoing an exponential expansion. Regardless, analytic corrections for the effect of genetic draft have yet to be derived, and would provide for a more straightforward way of accounting for this confounding factor.

Our results provide a simple analytical correction to $\alpha_{MK}$ for the case of tightly linked sections of the genome that accounts for genetic draft. First, we note that the number of polymorphic sites is closely related to heterozygosity $\pi$, the average number of nucleotide differences between two randomly drawn individuals, through
\[ \pi = \langle \sum_{i=1}^p 2 x_i (1-x_i) \rangle = 2 p (\langle x \rangle - \langle x^2 \rangle).  \]
Thus, the ratio $p_n/p_s$ is simply
\[ \frac{p_n}{p_s} = \frac{\pi_n ( \langle x \rangle_s - \langle x^2 \rangle_s)}{\pi_s  (\langle x \rangle_n - \langle x^2 \rangle_n) } \approx \frac{\pi_n}{\pi_s}. \]
In general, the moments of the frequencies of synonymous and nonsynonymous sites may be measured straightforwardly from the site frequency spectrum of the sample. However, as we have already demonstrated, in the rapid adaptation limit the SFS of neutral and beneficial mutations is similar in form. Thus, the heterozygosities of beneficial and neutral mutations are largely determined by different numbers of neutral and selected sites --- i.e. by $p_n$ and $p_s$ --- rather than significantly different frequency distributions.

The heterozygosity for beneficial mutations is calculated from the moments of $\rho(x_k|x_0)$ (using the same method used in the calculation of the beneficial SFS) to be $ \pi \approx 2(q-1)$, with higher order corrections given in \citet{DWF}. The neutral heterozygosity is simply $2 U_n T_2$, where $T_2$ is the expected coalescence time for two randomly chosen individuals. The rate of beneficial substitutions is $1/ \langle \tau_q \rangle$, and the rate of neutral substitutions is $U_n$. Thus, given the rate of synonymous mutations, $U_{n,s}$, and the rate of neutral nonsynonymous mutations, $U_{n,n}$, the expected, measured value of $\alpha_{MK}$ (by naively plugging in each measured value in \eqref{MKtest} ) will be
\[ \langle \alpha_{MK, meas} \rangle = 1 - \frac{d_s}{d_n} \frac{p_n}{p_s}  \approx   1 - \frac{d_s}{d_n} \frac{\pi_n}{\pi_s}  = 1 - \left( \frac{U_{n,s}}{1/\langle \tau_q \rangle + U_{n,n}} \right) \left( \frac{2 T_2 U_{n,n}+2 (q-1)}{2 T_2 U_{n,s}} \right). \]
Clearly, only the first term of $\pi_n$ corresponds to nonadaptive sites. Thus, we have
\[ \langle \alpha_{MK, meas} \rangle = \frac{d_+}{d_n} - \left( \frac{U_{n,s}}{1/\langle \tau_q \rangle + U_{n,n}} \right) \left( \frac{2 (q-1)}{2 T_2 U_{n,s}} \right) = \frac{d_+}{d_n} - \frac{2 (q-1) d_s}{d_n \pi_s}. \]
This gives for the fraction of adaptive substitutions
\[ \frac{d_+}{d_n} = 1 - \frac{d_s}{d_n} \frac{p_n}{p_s} +  \frac{d_s}{d_n} \frac{2(q-1)}{\pi_s} = \langle \alpha_{MK, meas} \rangle  + \frac{d_s}{d_n} \frac{2(q-1)}{\pi_s}.  \]
In practice, the parameter $q$ is measurable from estimates of $N$, $s$, and $U_b$ or the distribution of fitnesses within the population.

The interpretation of this correction is intuitively simple. In populations where adaptation is rapid, the assumption that beneficial mutations do not contribute significantly to measured polymorphism breaks down. As a result, ascribing all measured nonsynonymous polymorphism $p_n$ to neutral (or deleterious) mutations results in an underestimate of $d_+/d_n$. In fact, in the limit of infinitely rapid adaptation, $q \rightarrow \infty$, $\alpha_{MK, meas} \rightarrow 1/2$, in contrast to the \textit{true} fraction of adaptive substitutions, $d_+/d_n \rightarrow 1$. Our model, then, provides a simple correction for this underestimate by predicting the expected fraction of observed polymorphism (relative to synonymous polymorphism) that arises from beneficial mutations.

\subsection{Applications to Forks and More Complicated Fitness Landscapes}
%the findings of the previous section are relevant for rugged fitness landscapes

Finally, because of the flexibility of our model with respect to the underlying fitness landscape, it is straightforward to generalize our results to a small class of more complex interactions between sites. Specifically, our findings readily generalize to the situation in which the set of beneficial mutations $\mathcal{B}$ may be partitioned into disjoint subsets $B_i$, $\mathcal{B}  = \bigcup_{i=1}^n B_i$, with corresponding mutation rates $\{\mu_1, ..., \mu_n\}$, such that an individual carrying a mutation from a subset $B_i$ cannot generate successful progeny that also carries a mutation from a subset $B_j \neq B_i$. Although this condition sounds quite restrictive, it bears reminding that mutations destined to fix must be near the high fitness nose of the wave when they first occur. Thus, this condition will be fulfilled if there is sign epistasis between these subsets that is at least as strong as $\sim s$. Heuristically, this formalism describes a population evolving on a fitness landscape exhibiting two or more disjoint evolutionary pathways.

For example, we may consider first the case of a forked fitness landscape, in which two evolutionary trajectories are available. We may assume that the lead of the population has just reached the crossing of the fork, and further mutations occur along one evolutionary trajectory with rate $\xi U_b$ and along the other with rate $(1- \xi)U_b$.\footnote{Similarly, if we are interested in the success of one particular evolutionary trajectory out of many, we can simply consider our trajectory of interest to comprise $\xi U_b$ of the total mutation rate, and the combination of the other trajectories to comprise $(1- \xi)U_b$.} Note that it makes no difference whether the mutation rate down one evolutionary pathway represents one possible mutation occurring at rate $\xi U_b$, or many possible mutations occurring at this total rate. In either case, the factor $\xi$ replaces the lineage fraction $x_0$ in the generating function for the next establishing class. Considering mutations into one path in the fork, the generating function for their contribution $\nu_1$ in the next fitness class is
\[ \langle e^{- \nu_1 z} \rangle = e^{-\xi z^{\alpha}}. \]
If the mutation rate past the first step returns to $U_b$ for both trajectories, then results for jump probabilities and sojourn times follow exactly as before, with $x_0 = \xi$. If mutation rates on the two trajectories remain at $\xi U_b$ and $(1-\xi)U_b$, then the generating function at each new step is modified straightforwardly according to
\[\langle e^{- \nu_{k+1} z} \rangle = \langle e^{-\nu_k \xi z^{\alpha}} \rangle \]
with $\nu_k$ the contribution of the trajectory in the previous step. All our results above can then be modified by including the corresponding factors of $\xi$ and $(1- \xi)$ at each step.

In short, genetic draft acts at each jump forward according to the amount a particular lineage, a particular site, or a particular evolutionary trajectory comprises the mutation rate into the next step. Thus, with minor modifications, our method can be applied to answer questions about the fates of all three, for a variety of different fitness landscapes.

These considerations provide a different perspective on the fates of populations evolving on rugged fitness landscapes, and particularly on the effect of a larger population size in avoiding local fitness peaks. Previous works have suggested that over certain timescales, smaller populations may have an advantage in adapting on these rugged landscapes, because their trajectories are more heterogeneous, whereas larger populations have an increased tendency to get stuck on local fitness peaks \citep{Szendro1, Jain1, Rozen1, HandelRozen}. However, our analysis suggests that if the landscape is dominated by several distinct uphill trajectories featuring mutational steps of similar size, large populations may be capable of travelling for many steps down multiple paths, effectively exploring the surrounding landscape before settling upon one particular uphill trajectory. For example, our analysis has shown that in the case of a simple fork with equal mutation rates down both pathways, a large, rapidly adapting population will typically explore about $2 q$ mutational steps forward before a particular pathway is closed off. The transition between this behavior and that considered in the work cited above evidently occurs between classical clonal interference, in which adaptation is dominated by the rare emergence of extremely fit mutants, and the multiple mutations regime, in which most fixed mutations are of roughly the same size. This, in turn, strongly depends on the distribution of fitness effects of mutations, with long-tailed or short-tailed distributions giving rise to dynamics dominated by clonal interference or multiple mutations, respectively \cite{DF, Fogle}. The different outcomes predicted by these two regimes could explain the lack of experimental consensus on the effect of population size on outcomes of adaptation \citep{Rozen1, Miller, Schoustra}.

\subsection{Conclusions and Future Work}

By using a simple model, we have made considerable headway in understanding how genetic draft affects the frequencies of mutations through a series of stochastic jumps, how these jumps affect genetic diversity, sojourn times and fixation times of mutations, and why these statistics resemble those derived from the Bolthausen-Sznitman coalescent. We then showed how our method leads to a simple correction to the McDonald-Kreitman test that accounts for linkage between beneficial mutations. Finally, we discussed how our analysis might be extended to describe evolution on certain classes of rugged fitness landscapes, which --- although admittedly very simple ---  nonetheless describe limiting behavior for sign epistasis between multiple evolutionary pathways.

%DWF 12 claims some universal coalescent behavior for a DFE, but I don't really understand this.

Still, our model has some shortcomings. First, recombination is neglected in this model, which makes our results applicable only to the evolution of microbial populations and tightly linked regions of the genomes of sexually reproducing organisms. Naturally, in cases where recombination is no longer rare, the effects of genetic draft are tempered as competing beneficial mutations recombine onto a single genetic background. Fitness classes that evolve disjointly in the asexual model are then allowed to mingle at each reproductive step, meaning that a series of stochastic jumps between classes no longer correctly describe the dynamics. Fortunately, the recent work of \citet{Neher1}, which accounts for the effects of occasional (facultative) outcrossing of clones (but not for the effect of newly arising mutations on the clonal background) provides a framework for combining these two sources of genetic draft. In particular, since common mutations must at one point propagate near the most-fit class, evolutionary dynamics in these populations are still largely informed by the distribution of haplotypes in the nose. This distribution would then obtain contributions both from mutations from the adjacent class and recombined haplotypes obtained from mating between less fit clones.

We also make use of the assumption of a single selection coefficient. Indeed, two facets of our model that are key in deriving analytical results --- the organization of clones according to fitness classes and the asymptotic freezing of frequencies in each class --- both break down when a single selection coefficient is replaced with some distribution of fitness effects. However, several works \citep{Good1, DF} have shown that even in populations with a distribution of fitness effects, evolutionary dynamics are well described by the use of an effective, or \textit{predominant} selection coefficient, which exactly coincides with the most common fixed mutational effect.
%Furthermore, this problem of a distribution of fitness effects may be amenable to a `coarse graining' approach of the kind used in recent work \citep{Good2}, whereby lineages of similar absolute fitnesses are grouped together into effective fitness classes, organized according to absolute fitness instead of the amount of beneficial mutations.
Still, the inclusion of a distribution of fitness effects, and the resulting unified understanding of the effects of multiple mutations and mutations of varying effect sizes in driving evolutionary dynamics, remains a promising subject of future work.

%this formula, more than Haldane's formula, will be relevant in real populations

%Furthermore, when considering the fates of polymorphic sites measured in a given population sample $n \ll N$, the effect of genetic draft is likely to be a stronger evolutionary force than drift for non-singleton sites. Any such site is almost certainly at a large enough frequency in the population to have already escaped drift, the likelihood of which is given by Haldane's formula, $P_{esc} \sim s$. The success or failure of the mutation is then entirely dependent on its ability to escape the effects of genetic draft, which at long times we have argued is simply $P_{esc} \sim x$, where $x$ is the frequency of the mutation in the population. The effect of draft is thus similar to drift in that it stochastically amplifies the fluctuations in the trajectories of mutations, albeit the time and lengthscale of these fluctuations are much greater than for drift.

%%%%%%%%%%%%%%%%%%%%%%%%%%%%%%%%%%%%%%%%%%%%%%%%
%%%%%%%%%%%%%%%%%%%%%%%%%%%%%%%%%%%%%%%%%%%%%%%%
%%%%%%%%%%%%%%%%%%%%%%%%%%%%%%%%%%%%%%%%%%%%%%%%
%%%%%%%%%%%%%% ACKNOWLEDGEMENTS %%%%%%%%%%%%%%%%%%%%%%%%%%%%
%%%%%%%%%%%%%%%%%%%%%%%%%%%%%%%%%%%%%%%%%%%%%%%%
%%%%%%%%%%%%%%%%%%%%%%%%%%%%%%%%%%%%%%%%%%%%%%%%

\section{Acknowledgements}
 We would like to thank Benjamin Good and Sergey Kryazhimskiy for their helpful comments and suggestions. This work was supported by a National Science Foundation Graduate Research Fellowship (K.K.) and by the James S. McDonnell Foundation and the Alfred P. Sloan Foundation (M.M.D.). %Simulations were performed on the Odyssey cluster supported by the Research Computing Group at Harvard University.

%%%%%%%%%%%%%%%%%%%%%%%%%%%%%%%%%%%%%%%%%%%%%%%%
%%%%%%%%%%%%%%%%%%%%%%%%%%%%%%%%%%%%%%%%%%%%%%%%
%%%%%%%%%%%%%%%%%%%%%%%%%%%%%%%%%%%%%%%%%%%%%%%%
%%%%%%%%%%%%%% APPENDICES %%%%%%%%%%%%%%%%%%%%%%%%%%%%
%%%%%%%%%%%%%%%%%%%%%%%%%%%%%%%%%%%%%%%%%%%%%%%%
%%%%%%%%%%%%%%%%%%%%%%%%%%%%%%%%%%%%%%%%%%%%%%%%

\begin{appendices}

%%%%%%%%%%%%%%%%%%%%%%%%%%%%%%%%%%%%%%%%%%%%%%%%%%%%%%%%%
%%%%%%%%%%%%%%%%%%% APPENDIX  A %%%%%%%%%%%%%%%%%%%%%%%%%%%%%%%%
%%%%%%%%%%%%%%%%%%%%%%%%%%%%%%%%%%%%%%%%%%%%%%%%%%%%%%%%%
\section{Appendix A: Dynamics of the Transition Process}
\manuallabel{transitionprobs}{A}

 In this appendix we derive the justification that frequencies of mutant lineages are frozen when a class begins feeding mutants to the lead that are destined to establish. We then explicitly derive the probability distribution $\rho_k$ of a transition in a mutation's frequency from a starting class $0$ to some final fitness class $k$.

 To prove that frequencies of mutant lineages are frozen when a class begins supplying establishing mutants to the next class, we first note that the $L$-th establishing mutant in a given fitness class typically occurs at time $t_L$ such that
\[ L = \underbrace{qs}_{(a)}\cdot  \underbrace{U_b \int_0^{t_L} \frac{e^{(q-1) s t}}{qs} dt}_{(b)}, \]
where $(a)$ is the establishment probability of one mutant and $(b)$ is the total number of mutants introduced into the lead class by time $t_L$. Thus, using the same argument as in \citet{DWF}, the amount that the $L$-th establishing lineage contributes to a fitness class as a fraction of the first lineage is
\[ \frac{\eta_{L,k}(t)}{\eta_{1,k}(t)} = \frac{ e^{q s (t-t_L)}}{e^{q s (t-t_1)}} = \frac{1}{L^{q/(q-1)}}. \]
Note that this is an \textit{upper} limit on the contribution of $\eta_{L,k}$, since the growth of each subsequent lineage actually decreases according to the rate of adaptation $v$. A fitness class typically  establishes in a time $\langle \tau_q \rangle$, and generates its first establishing mutant a time $t_1$ after that. If we neglect the decreasing growth rate due to adaptation of the population (valid for large $q$), then at this point, the class below it has supplied
\[ qs U_b \int_0^{t_1 + \langle \tau_q \rangle}  \frac{e^{(q-1) s t}}{qs} dt \sim \frac{s}{U_b}  \gg 1 \]
establishing mutants. 

In practice, however, few biological populations evolve with $q>4$, in which case the diminishing growth rate becomes important in the above calculation. A simple modification of the analysis gives the number of establishing lineages to be roughly $$\left( (q-2)(q-1)^{(q-3)/(q-1)}/((q-3)q) \right) \left(s/U_b \right)^{(q-2)/(q-1)}$$ for $q>3$, which, although considerably smaller than the asymptotic value of $s/U_b$ is offset by the more rapid decay of $\eta_{L,k}/\eta_{1,k}$ for smaller $q$. Extensions for $q=2,3$ are likewise straightforward. 

Thus, we have demonstrated that the contribution of subsequent lineages diminishes rapidly, and by the time a fitness class begins feeding establishing mutants to the class below it, it has $\mathcal{O}(s/U_b) \gg 1$ lineages that are destined to establish in it. As a result, the contribution of subsequent lineages after this time is already very small, meaning that the frequencies of common lineages in the fitness class at this time may safely be treated as frozen.

The astute reader might also note that mutant lineages that are destined to go extinct may also contribute to shifting the frequencies, which is certainly a contributing factor close to the time that the class establishes. Although not as straightforward a calculation, if $s \gg U_b$ then it is still true that by the time a class begins feeding establishing mutations, the contribution of these ``doomed lineages'' is also small (Appendix \ref{doomedappendix}). Thus we are justified in treating the process as a Markov chain. %, with transition probabilities of lineages from class to class given by $\rho(f_l(k)| f_l(k-1))$.

%Using these assumptions, given a lineage and its complement, we would like to find the probability distribution given by BLAH, given the generating function nu_l, nu^{\prime}_l. The probability of a transition is BLAH (Appendix B)

Now that we have justified our Markov chain assumption, we are ready to derive the transition probability of a mutation at frequency $x_0$ in class $0$ to some frequency $x_{k}$ in class $k$. It has been shown \citep{DF} that the contribution to class $k$, $\nu_k$, from a lineage at frequency $x_{k-1}$ in class $k-1$ is described by the generating function
\[ \langle e^{-\nu_k z} \rangle = e^{-x_{k-1} z^{\alpha}}. \]
Thus, after class $k$ establishes, the lineage grows as
\[ \eta_k(t) =  \frac{x_k}{qs} e^{(q-1) s (t - \tau_q )} \propto \frac{\nu_k }{qs} e^{(q-1) s (t - \langle \tau_q \rangle)}. \]

The generating function for the contribution $\nu_{k+1}$ of the mutation $2$ fitness steps forward is then obtained by averaging over the intermediate value $\nu_k:$
\begin{align}
 \langle e^{-\nu_{k+1} z} \rangle &= \langle e^{-\nu_{k} z^{\alpha}} \rangle_{\nu_{k}} = \int_0^{\infty}  e^{-\nu_k z^{\alpha}} P(\nu_k| x_{k-1}) d \nu_k  \nonumber \\
&= \frac{1}{2 \pi i} \int_0^{\infty} \int_{ - i \infty}^{ i \infty} e^{-\nu_k z^{\alpha}} e^{\nu_k \zeta} e^{-x_{k-1} \zeta^{\alpha}} d\zeta d \nu_k   \nonumber \\
&= e^{- x_{k-1} z^{\alpha^2}}, \nonumber
\end{align}
where the step in line two makes use of the standard formula for inverting the Laplace transform. Although technically the average should be taken over the frequency $x_k$ (which is analytically intractable), an average over $\nu_k$ is a reasonable approximation if $\sigma$, the growth of the entire class, does not deviate too much from its typical value, and if the growth of the lineage $\nu_k$ in future classes may be taken independently of the growth of other lineages. Because the effect of this approximation is compounded at each step forward, it introduces significant deviations at timescales of roughly $q$ fitness steps forward, which is the timescale over which fluctuations in the advance of the fitness wave become significant. Accepting this approximation, it is then straightforward to show that
\[ \langle e^{- \nu_{k} z} \rangle = e^{- x_0 z^{\alpha^k}}. \]
\indent Without loss of generality, we can index the class in which the mutant begins to be tracked to 0. The fitness class at $k=1$ can then be divided into those individuals descended from a particular lineage in class $0$, at frequency $x_0$, and those not descended from that lineage, at frequency $(1-x_0)$. Since, for each subsequent fitness class, new frequencies are frozen near the nose, where the two sets of individuals proliferate independently, there are two independent variables encoding the fate of the lineage: $\nu_k$, denoting the contribution of the lineage to a class with $k$ more beneficial mutations, and $\tilde{\nu}_k$, denoting the contribution of individuals not derived from that lineage, so that $\tilde{\nu}_k + \nu_k = \sigma_k$.

If we denote the probability densities of $\nu_k$ and $\tilde{x}_k = \tilde{\nu}_k/\nu_k$ by $P_k(\nu_k| x_0)$ and $\Pi_k(\tilde{x}_k| x_0)$ respectively, then
\begin{align}
 \Pi_k(\tilde{x}_k| x_0) &= \int_0^{\infty} P_k(\tilde{x}_k \nu_k| 1- x_0) P_k(\nu_k| x_0) \nu_k d \nu_k \nonumber \\
&= \frac{1}{(2 \pi i)^2} \int_0^{\infty} \int_{\epsilon - i \infty}^{\epsilon + i \infty} \int_{ - i \infty}^{ i \infty}   e^{\tilde{x}_k \nu_k z_1} e^{\nu_k  z_2} e^{-(1- x_0) z_1^{\alpha^k}} e^{- x_0 z_2^{\alpha^k}} \nu_k dz_1 dz_2 d \nu_k, \nonumber
\end{align}
where $\epsilon \in \mathbb{R}, \epsilon \rightarrow 0^- $. Evaluating all the integrals gives
\begin{equation} \Pi_k(\tilde{x}_k| x_0) =  \frac{ \sin(\pi \alpha^k)}{\tilde{x}_k \pi} \frac{ x_0(1- x_0)}{(1- x_0)^2 \tilde{x}_k^{-\alpha^k} + x_0^2  \tilde{x}_k^{\alpha^k} +  2 x_0 (1- x_0)\cos(\pi \alpha^k) }. \label{Pi} \end{equation}
Finally, we perform a change of variables to $x_k = \frac{1}{1+\tilde{x}_k}$, giving the jump distribution,
%The transition probability is then given by the probability distribution of $x_k = \nu_k/\sigma$: (derived in Appendix \label{transitionprobability})
\begin{equation} \rho_k(x_k | x_0) = \frac{ \sin(\pi \alpha^k) x_0(1- x_0)}{x_k (1-x_k) \pi \left[(1- x_0)^2 \left(\frac{x_k}{1-x_k}\right)^{\alpha^k} +  x_0^2 \left(\frac{1-x_k}{x_k}\right)^{\alpha^k} +  2   x_0 (1-  x_0)\cos(\pi \alpha^k) \right] }. \end{equation}
$\rho_k$ denotes the probability density of observing a mutation at frequency $x_k$ in class $k$, given that it was at frequency $x_0$ in class $0$. The function cited in the text is $\rho(x_1| x_0) = \rho_1(x_1|x_0)$.
%%%%%%%%%%%%%%%%%%%%%%%%%%%%%%%%%%%%%%%%%%%%%%%%%%%%%%%%%
%%%%%%%%%%%%%%%%%%% APPENDIX  B %%%%%%%%%%%%%%%%%%%%%%%%%%%%%%%%
%%%%%%%%%%%%%%%%%%%%%%%%%%%%%%%%%%%%%%%%%%%%%%%%%%%%%%%%%

\section{Appendix B: Contribution of doomed lineages to a fitness class}
\manuallabel{doomedappendix}{B}

In deriving the transition probabilities, we argue that the frequency of competing mutant lineages is frozen by the time a fitness class begins feeding mutants that are destined to establish into the nose. We arrive at this conclusion by calculating the number of establishing mutant lineages by the (typical) time this occurs. Since this number is typically large, and the contribution of each subsequent establishing lineage vanishes (despite the fact that these lineages arrive more and more quickly), the conclusion follows that the frequencies of lineages by this timepoint are frozen. This comes from the fact that the deterministic lineage size fraction, $n_L /n_1 \sim 1/L^{q/(q-1)}$ is convergent.

What could be the contribution of mutants that will not establish in a class, but may still contribute establishing mutants to the \textit{next} fitness class? There are many more mutants introduced that will not establish relative to those that will, and these mutants grow roughly proportionally to the size of the class below them (i.e., at rate $(q-1)s$). By the time the population starts supplying establishing mutants, if these doomed lineages still comprise a significant fraction of the population, the result will be a deterministic drifting of the frequency, as these doomed lineages (which, because they are plentiful, will typically be split as $x_0$) become less and less of a contributing factor. We would like to determine if these combined lineages are negligible by the time a class begins feeding establishing mutants or still constituting some non-trivial fraction of the class.

The total number of individuals at time $t$ derived from lineages that are destined to go extinct is given by
\[ N_{doomed}(t) = \int_0^t \underbrace{d \tau U_b (1-qs) \frac{1}{qs} e^{(q-1) s \tau}}_{(a)} \underbrace{ \langle n (t-\tau) | n(t \rightarrow \infty) = 0 \rangle}_{(b)}, \]
where $(a)$ is the expected number of non-establishing mutants occurring in a small interval $d\tau$, and $(b)$ is the expected number of these mutants that still persist at a time $t-\tau$ later, given that these mutants are going to eventually go extinct. Here, $\langle n(t- \tau)|n(t \rightarrow \infty) = 0  \rangle$ is derived from standard branching process analyses.

The probability distribution of doomed (but not yet extinct) lineages at time $t$ is given by
\begin{equation}
 P(n>0,t | \text{doomed}) = P_{ext}(n) P(n,t)/P_{ext}(t), \label{pnt}
\end{equation}
where $P_{ext}(n)$ is the probability of extinction of a lineage composed of $n$ individuals, $P_{ext}(t)$ is the probability that a lineage destined to go extinct is not yet extinct by time $t$, and $P(n,t)$ is the probability that a single mutant created at $t=0$ has $n$ descendants at time $t$. Note that by a ``doomed'' lineage we mean a lineage that is destined to vanish.

%\[ P(n,t | \text{doomed}) = \frac{P_{ext}(n) P(n,t)}{ \sum_n P_{ext}(n) P(n,t)} \]
%
%where $P_{ext}(n)$ is the probability of extinction of a lineage composed of $n$ individuals and $P(n,t)$ is the probability that a single mutant created at $t=0$ has $n$ ancestors at time $t$.
%
Now, following standard branching process analysis \citep{DF}, a new mutant at fitness $qs$ will have the following distribution of ancestors a time $t$ later:
\[ P(n=0, t) =  \frac{ e^{qst}-1}{(1+qs)e^{qst}-1}, \]
so that
\[P(n=0, t | \text{doomed}) = \frac{(1+qs)( e^{qst}-1)}{(1+qs)e^{qst}-1}, \]
and furthermore
\[ P(n>0, t) = \frac{(qs)^2 e^{qst}}{((1+qs)e^{qst}-1)((1+qs)e^{qst}-1-qs)} \left( \frac{((1+qs)e^{qst}-1-qs)}{(1+qs)e^{qst}-1} \right)^n. \]
In the branching process analysis, all lineages are independent, so
\[ P_{ext}(n)= P_{ext}(1)^n = \left( \frac{1}{1+q s} \right)^n. \]
Furthermore,
\[ P_{ext}(t) = 1 - P(n=0, t | \text{doomed}) = \frac{qs}{(1+qs) e^{qst}-1}. \]
%\begin{align}
%  P(n>0,t | \text{doomed}) &= \left( \frac{e^{qst}-1}{(1+qs)e^{qst}-1} \right)^n \left( \sum_n \left( \frac{e^{qst}-1}{(1+qs)e^{qst}-1} \right)^n \right)^{-1} \nonumber \\
%&=  \left( \frac{e^{qst}-1}{(1+qs)e^{qst}-1} \right)^n \frac{(1+qs) e^{qst}-1}{qs e^{qst}} \nonumber \\
%\end{align}
Thus, using \eqref{pnt} gives
\begin{align}
P(n>0,t | \text{doomed})= \frac{q s e^{qst}}{(1+qs)e^{qst}-1-qs}\left[ \frac{e^{qst}-1}{(1+qs)e^{qst}-1} \right]^n . \nonumber
\end{align}
Finally, this gives us
\begin{align}
 \langle n (t) | n(t) \neq 0, n(t \rightarrow \infty) = 0 \rangle &=  \sum_{n=1}^{\infty}  n P(n,t | \text{doomed}) \nonumber \\
&=  \frac{q s e^{qst}}{(1+qs)e^{qst}-1-qs}  \sum_{n=1}^{\infty}  n \xi^n \nonumber
\end{align}
with \[ \xi = \frac{e^{qst}-1}{(1+qs)e^{qst}-1}. \] We then use the following identity:
\begin{align}
 \sum_{n=0}^{\infty} n  \xi^n &= \frac{\xi}{(1-\xi)^2}. \nonumber
\end{align}
As a result, $ \langle n (t) | n(t) \neq 0, n(t \rightarrow \infty) = 0 \rangle$ reduces to
\begin{align}
  \langle n (t) | n(t) \neq 0,  n(t \rightarrow \infty) = \frac{((1+qs)- e^{-qst})}{qs} \nonumber
\end{align}
Now, we include the possibility that the lineage has gone extinct by time $t$:
\begin{align}
\langle n (t)|  n(t \rightarrow \infty) = 0 \rangle &=  \frac{((1+qs)- e^{-qst})}{qs} P[n \neq 0 | \text{doomed}] \nonumber \\
&= e^{-qst}. \nonumber
\end{align}
%This reduces to the correct limit at $t=0, \infty$.
\indent A fitness class, naively, has $\mathcal{O}(1/U_b)$ individuals when it typically starts creating establishing mutants. Hence,
\begin{align}
 \frac{N_{doomed}(t_1+\tau_q)}{n_k((t_1+\tau_q))} &= U_b^2 \int_0^{(t_1+\tau_q)} (1-qs) \frac{1}{qs} e^{(q-1) s \tau} e^{-qs((t_1+\tau_q)-\tau)} d\tau  \nonumber \\
&\approx \frac{U_b^2 e^{qs (t_1 + \tau_q)}}{2 q s} \nonumber \\
&\approx \frac{s}{2}. \nonumber
\end{align}
Thus, given our assumption that $s \ll 1$, the contribution of doomed lineages is indeed small and can be ignored.
%%%%%%%%%%%%%%%%%%%%%%%%%%%%%%%%%%%%%%%%%%%%%%%%%%%%%%%%%
%%%%%%%%%%%%%%%%%%% APPENDIX C %%%%%%%%%%%%%%%%%%%%%%%%%%%%%%%%
%%%%%%%%%%%%%%%%%%%%%%%%%%%%%%%%%%%%%%%%%%%%%%%%%%%%%%%%%
\section{Appendix C: Moments of the jump probability}
\manuallabel{moments}{C}
The moments of the jump probability $\rho$ are calculated as follows, analogous to the method used in \citet{DWF}. Using the fact that
\[ \left( \frac{1}{\sigma} \right)^n = \int_0^{\infty} \frac{z^{n-1}}{(n-1)!} e^{-z \sigma} dz, \]
the $n$-th moment (for the step $k \rightarrow k+1$) is derived as
\begin{align}
 \left\langle x^n \right\rangle = \left\langle \left( \frac{\nu}{\sigma} \right)^n \right\rangle &= \left\langle \int_0^{\infty} \frac{z^{n-1}}{(n-1)!} e^{-z \sigma} \nu^n dz \right\rangle \nonumber \\
&=   \int_0^{\infty} \frac{z^{n-1}}{(n-1)!} \left\langle e^{-z \sigma} \nu^n  \right\rangle dz \nonumber \\
&=   \int_0^{\infty} \frac{z^{n-1}}{(n-1)!} \left\langle e^{-z (\nu+\tilde{\nu})} \nu^n  \right\rangle dz \nonumber \\
&=   \int_0^{\infty} \frac{z^{n-1}}{(n-1)!} \left\langle e^{-z \tilde{\nu}} \right\rangle \left\langle e^{-z \nu} \nu^n  \right\rangle dz \nonumber \\
&= \int_0^{\infty} \frac{z^{n-1}}{(n-1)!} e^{-(1-x_0)z^{(1-1/q)}} \left\langle e^{-z \nu} \nu^n  \right\rangle dz. \nonumber
\end{align}
\indent Now,
\[ \langle \nu^n e^{-z \nu} \rangle = (-1)^n \frac{d^n}{dz^n} \langle e^{-z \nu} \rangle \]
and
\[  \langle e^{-z \nu} \rangle = e^{- x_0 z^{(1-1/q)}}. \]
We can thus compute a general formula for the $n$-th derivative:
\[ \frac{d^n}{dz^n} e^{-x_0 z^{(1-1/q)}} = e^{-x_0 z^{(1-1/q)}} \sum_{i=0}^{n-1} (-x_0)^{n-i} (1-1/q)^{n-i} z^{-(n-i)/q - i} (-1)^{i} \frac{\Gamma(1/q+i)}{\Gamma(1/q)}. \]
Thus,
\[ \langle \nu^n e^{-z \nu} \rangle = e^{-x_0 z^{1-1/q}} \sum_{i=0}^{n-1} x_0^{n-i} (1-1/q)^{n-i} z^{-(n-i)/q - i} \frac{\Gamma(1/q+i)}{\Gamma(1/q)}. \]
This gives for the moments,
\begin{align}
 \left\langle x^n \right\rangle = \left\langle \left( \frac{\nu}{\sigma} \right)^n \right\rangle &=  \int_0^{\infty} \frac{z^{(n-1)}}{(n-1)!} e^{-(1-x_0)z^{1-1/q}} e^{-x_0 z^{1-1/q}} \sum_{i=0}^{n-1} x_0^{n-i} (1-1/q)^{n-i} z^{-(n-i)/q - i} \frac{\Gamma(1/q+i)}{\Gamma(1/q)} dz \nonumber \\
&= \sum_{i=0}^{n-1} x_0^{n-i} (1-1/q)^{n-i} \frac{\Gamma(1/q+i)}{(n-1)! \Gamma(1/q)} \int_0^{\infty} z^{(n-1)} e^{-z^{1-1/q}}z^{-(n-i)/q - i}  dz \nonumber \\
&= \sum_{i=0}^{n-1} x_0^{n-i} \frac{\Gamma(1/q+i) \Gamma(n-i)}{(n-1)! \Gamma(1/q)} \left( \frac{q-1}{q} \right)^{n-1-i}.
\label{momentformula}
\end{align}
The first two moments are
\begin{align}
 \langle \Delta x \rangle &= 0, \nonumber \\
 \langle (\Delta x)^2 \rangle &= \frac{x_0(1-x_0)}{q}. \nonumber
\end{align}
Equation \eqref{momentformula} is readily generalized to $k$ steps forward through the usual substitution
\[ 1-1/q = \alpha \Rightarrow \alpha^k, \]
which gives
\[ \left\langle x_k^n \right\rangle =  \sum_{i=0}^{n-1} x_0^{n-i} \frac{\Gamma(1- \alpha^k+i) \Gamma(n-i)}{(n-1)! \Gamma(1-\alpha^k)} \left( \alpha^{k} \right)^{n-1-i}.  \]
%%%%%%%%%%%%%%%%%%%%%%%%%%%%%%%%%%%%%%%%%%%%%%%%%%%%%%%%%
%%%%%%%%%%%%%%%%%%% APPENDIX D %%%%%%%%%%%%%%%%%%%%%%%%%%%%%%%%
%%%%%%%%%%%%%%%%%%%%%%%%%%%%%%%%%%%%%%%%%%%%%%%%%%%%%%%%%
\section{Appendix D: Site Frequency Spectrum of Nearly Private Variants}
\manuallabel{rareSFSappendix}{D}

Here we derive the leading order behavior for the integral given in \eqref{frare}, describing the expected number of sites that are almost private (i.e., with lineage sizes $n \ll \sqrt{2/v}$).
\begin{align}
& F_{rare}(n/N,y) = \frac{1}{v} \int_{y}^{(q-2)s} P_1 (n, (y_0 - y)/v) U_b N(y_0 - s) d x_0 \nonumber \\
&=  \frac{1}{v} \frac{U_b N}{\sqrt{2 \pi v}} \int_{y}^{\infty}\frac{ e^{\frac{  y^2+ 2 y_0 s-2 y_0^2- s^2}{2 v}}}{(\int_0^{((y_0 - y)/v)} e^{- y_0 \tau + v \tau^2/2}d\tau+1)^2}  \left( 1- \frac{ e^{ \frac{  y^2 - y_0^2}{2 v}}}{\int_0^{((y_0 - y)/v)} e^{- y_0 \tau + v \tau^2/2}d\tau+1}  \right)^{n-1} d y_0. \nonumber
\end{align}
The population-wide, rare variant SFS is obtained by integrating over all final fitnesses:
\begin{align}
F_{rare}(n/N) =& \frac{1}{v} \frac{U_b N}{\sqrt{2 \pi v}} \int_{-\infty}^{\infty} \int_{y}^{\infty}\frac{ e^{\frac{  y^2+ 2 y_0 s-2 y_0^2- s^2}{2 v}}}{(\int_0^{((y_0 - y)/v)} e^{- y_0 \tau + v \tau^2/2}d\tau+1)^2}  \nonumber \\
&\times \left(1 - \frac{e^{ \frac{  y^2 - y_0^2}{2 v}}}{\int_0^{((y_0 - y)/v)} e^{- y_0 \tau + v \tau^2/2}d\tau+1}  \right)^{n-1} d y_0 dy. \nonumber
\end{align}
To extract the leading order behavior, we consider frequencies just at the mean, $y=0$:
\[ F_{rare}(n/N, 0) = \frac{1}{v} \frac{U_b N}{\sqrt{2 \pi v}} \int_{0}^{\infty}\frac{ e^{\frac{2 y_0 s-2 y_0^2- s^2}{2 v}}}{(\int_0^{y_0 /v} e^{- y_0 \tau + v \tau^2/2}d\tau+1)^2} \left(1 - \frac{e^{ \frac{- y_0^2}{2 v}}}{\int_0^{y_0/v} e^{- y_0 \tau + v \tau^2/2}d\tau+1}  \right)^{n-1} d y_0. \]
Examining the integral inside the integrand, we observe that
\begin{align}
 \int_0^{y_0/v} e^{- y_0 \tau + v \tau^2/2}d\tau &=  \sqrt{ \frac{2}{v}} e^{-y_1^2} \int_0^{y_1} e^{y^2} dy = \sqrt{ \frac{2}{v}}  D(y_1) \nonumber
\end{align}
for $y_1 = y_0/\sqrt{2 v}$, and $D(y_1)$ is Dawson's integral, a well-studied special function. If $y_1$ is small, then
\[ D(y_1) \approx y_1 - \frac{2}{3} y_1^3 + \frac{4}{15} y_1^5 ... \]
\indent Since we are considering only small lineage sizes $n$, the integral will obtain its main contribution for $y_0$ small (meaning that most nearly-private variants were founded recently in the past). Thus we can use the first order expansion:
\begin{align}
F_{rare}(n/N,0) & \approx  \frac{U_b N}{v \sqrt{2 \pi v}} \int_{0}^{\infty}\frac{ e^{\frac{2 y_0 s-2 y_0^2- s^2}{2 v}}}{ (y_0/v +1)^2}  \left( 1 - \frac{1}{ y_0/v+1}  \right)^{n-1} d y_0 \nonumber \\
&= \frac{U_b N v e^{-s^2/(2v)}}{\sqrt{2 \pi v}} \int_{0}^{\infty}  \exp \left( \frac{y_0 s-y_0^2}{v} \right) \frac{y_0^{n-1}}{(y_0+ v)^{n+1}} d y_0 \nonumber \\
& \approx  \frac{U_b N v e^{-s^2/(2v)}}{ \sqrt{2 \pi v}} \bigg( \int_{0}^{\infty}  \exp \left( \frac{- y_0^2}{v} \right) \frac{ y_0^{n-1}}{(y_0+ v)^{n+1}} d y_0 \nonumber \\
&+ \frac{s}{v}  \int_{0}^{\infty}  \exp \left( \frac{- y_0^2}{v} \right) \frac{y_0^{n}}{(y_0+ v)^{n+1}} d y_0  \bigg)  \nonumber \\
 &= \frac{U_b N e^{-s^2/(2v)}  }{\sqrt{2 \pi}} \bigg(  \int_{0}^{\infty}  \frac{ e^{-\xi^2} \xi^{n-1}}{( \xi+ \sqrt{v})^{n+1}} d \xi +  \frac{s}{\sqrt{v}}  \int_{0}^{\infty} \frac{e^{-\xi^2} \xi^{n}}{(\xi+\sqrt{v})^{n+1}} d \xi  \bigg),  \nonumber
\end{align}
where, in the last step, the substitution $\xi = \frac{y_0}{\sqrt{v}}$ was performed.
Performing each integration separately:
\begin{align}
\int_{0}^{\infty}  \frac{ e^{-\xi^2} \xi^{n-1}}{( \xi+ \sqrt{v})^{n+1}} d \xi &\approx  \frac{1}{n \sqrt{v}}, \nonumber \\
\int_{0}^{\infty}  \frac{ e^{-\xi^2} \xi^{n}}{( \xi + \sqrt{v})^{n+1}} d \xi &\approx  -\frac{\gamma}{2} + \log \left( \frac{1}{n \sqrt{v}}  \right) + \frac{1}{n} \nonumber
\end{align}
for $\sqrt{v} \ll 1$. Note that the second approximation breaks down for $n$ large; however, in the realm of validity of our approximation (derived below), the second term will generally be much smaller than the first term because of the log dependence on $n$.

Thus, the leading order behavior is
\[ F_{rare}(n/N, y=0) = \frac{U_b N e^{-s^2/(2v)}  }{\sqrt{2 \pi v} n}.\]

\subsection{Small $n$ approximation condition}
We would like to derive a realm of validity for all of our approximations. The primary assumption made in simplifying the Dawson's integral and exponential integrals is that $y_0/\sqrt{2v} \ll 1$. We observe that since we are integrating against a Gaussian, the integral is sharply peaked. In particular, the peak is dominated by the maximum of
\[ e^{\frac{2 y_0 s-2 y_0^2- s^2}{2 v}}  \exp \left( - \frac{(n-1) e^{ \frac{- y_0^2}{2 v}}}{ e^{-y_0^2/2v} \int_{0}^{y_0/v} e^{  v y^2 /2} dy+1}  \right). \]
This means that we would like to find the maximum of
\[ \frac{2 y_0 s-2 y_0^2- s^2}{2 v} - \frac{(n-1) e^{ \frac{- y_0^2}{2 v}}}{ e^{-y_0^2/2v} \int_{0}^{y_0/v} e^{  v y^2 /2} dy+1} . \]

Define
\[ I = e^{-y_0^2/2v} \int_{0}^{y_0/v} e^{  v y^2 /2} dy. \]
The equation to be solved for the peak $y_{max}$ is
\begin{align}
0 &= s - 2 y_{max}+ \frac{(n-1)y_{max}}{I+1} e^{-y_{max}^2/(2v)} + \frac{ (n-1)  y_{max} I e^{-y_{max}^2/(2v)}}{(I+1)^2}. \nonumber
\end{align}
Note that $y_{max} = \mathcal{O}(s/2)$, with some $n$ dependent correction that necessarily increases the location of the peak. Thus, for any of these Taylor expansions to hold for any $n$, we require at minimum that $s/2 \ll \sqrt{2 v}$.

In this case it is true that
\[ y_{max} = \frac{s}{2}+ \frac{(n-1)y_{max}}{2(I+1)} e^{-y_{max}^2/(2v)} + \frac{ (n-1)  y_{max} I e^{-y_{max}^2/(2v)}}{2(I+1)^2} <  \frac{s}{2}+ \frac{(n-1)y_{max}}{2(I+1)} + \frac{ (n-1)  y_{max} I}{2(I+1)^2} . \]
This may be solved numerically to find the exact location of the peak. However, if we suppose that the $n$ terms are a small perturbation on the $s/2$ peak, and (as we have already assumed) $s/2 \ll \sqrt{2v}$, we are justified in a first order expansion of Dawson's integral:
\begin{align}
 y_{max} &< \frac{s}{2}+ \frac{(n-1)y_{max}}{2(y_{max}/v+1)} + \frac{ (n-1)  y_{max}^2/v}{2(y_{max}/v+1)^2}   \nonumber \\
&<  \frac{s}{2}+ (n-1) v . \nonumber
\end{align}
So long as $(n-1) v \ll 1$, it is indeed true that the new maximum is a small perturbation around the $s/2$ peak, and our original first order expansion of $I$ was justified.

Thus, for the small $n$ approximation to hold, we require that
$y_{max} < \frac{s}{2}+(n-1) v \ll \sqrt{2v} .$
Thus, the small $n$ approximation holds for
\[ n \ll \sqrt{ \frac{2}{v}} - \frac{s}{2v} +1. \]
%%%%%%%%%%%%%%%%%%%%%%%%%%%%%%%%%%%%%%%%%%%%%%%%%%%%%%%%%
%%%%%%%%%%%%%%%%%%% APPENDIX E %%%%%%%%%%%%%%%%%%%%%%%%%%%%%%%%
%%%%%%%%%%%%%%%%%%%%%%%%%%%%%%%%%%%%%%%%%%%%%%%%%%%%%%%%%
\section{Appendix E: Derivation of the Neutral Site Frequency Spectrum}
\manuallabel{neutralsfsappendix}{E}

In this section we derive the asymptotic form of the neutral site frequency spectrum for common alleles. To derive this, we start from the assumption of a class that is growing exponentially (and deterministically), such that
\[ n_{k-1}(t) = \frac{e^{(q-1)s t}}{qs}.\]
This class supplies neutral mutants at a rate $U_n$. Thus the expected number of individuals with lineage sizes $n$ at time $t$, $F(n,t)$, is simply the integral over the expected number of mutants introduced at time $\tau$ multiplied by the probability for a mutant to reach a lineage size $n$ in time $t- \tau$:
\[ F(n, t) = \int_{-\infty}^t U_n n_{k-1}(\tau) P(n, t - \tau) d \tau.  \]

The probability for a lineage to reach size $n$ in a time $t$ is given in \citet{DF} and in Appendix \ref{doomedappendix}. This gives
\begin{align}
F(n,t) &= ((q-1)s)^2 \int_{-\infty}^t \frac{U_n d \tau}{qs} e^{(q-1)s \tau} \left( \frac{(1+(q-1)s)e^{(q-1)s(t- \tau)} -1 - (q-1)s)}{((1+(q-1)s)e^{(q-1)s(t- \tau)} -1)} \right)^n \nonumber \\
 & \times \frac{e^{(q-1)s(t- \tau)}}{((1+(q-1)s)e^{(q-1)s(t- \tau)} -1)((1+(q-1)s)e^{(q-1)s(t- \tau)} -1 - (q-1)s)} .\nonumber
 \end{align}
 Defining $\sigma  =(q-1)s, y = (1+\sigma) e^{\sigma k} - 1 - \sigma$, we obtain
 \begin{align}
  F (n,t) &=  \frac{\sigma U_n e^{\sigma t}}{qs} \int_{0}^{\infty}  \frac{ dy y^{n-1}}{(y+1+\sigma)(y+ \sigma)^{n+1}}. \nonumber
  \end{align}
\indent An explicit series expansion in powers of $n$ may now explicitly be derived. First we note that
\[ \frac{1}{y+1+\sigma} = \frac{1}{y+\sigma} \left( \frac{1}{1+ \frac{1}{y+\sigma}} \right) =  \sum_{k=0}^{\infty} \frac{(-1)^k}{(y+\sigma)^{k+1}},  \]
which gives
\begin{align}
  F(n,t)  &=  \frac{\sigma U_n e^{\sigma t}}{qs} \int_{0}^{\infty} \sum_{k=0}^{\infty} \frac{ dy y^{n-1} (-1)^k}{(y+ \sigma)^{n+k+1}} = \frac{U_n e^{\sigma t}}{qs} \sum_{k=2}^{\infty} \frac{(-1)^k (k-1)! \Gamma(n)}{\sigma^{k-1} \Gamma(k+n)} =  \frac{U_n e^{\sigma t}}{qs} \sum_{k=2}^{\infty} \frac{(-1)^k \beta(k,n)}{\sigma^{k-1}},  \nonumber
  \end{align}
  where the Beta function $\beta(k,n)$ is defined as
  \[ \beta(k,n) = \frac{\Gamma(k) \Gamma(n)}{\Gamma(n+k)}. \]
\indent The above is well approximated by the following expansion:
\[  F(n,t) =  \frac{U_n e^{\sigma t}}{qs} \sum_{k=2}^{\infty} \frac{(-1)^k (k-1)!}{\sigma^{k-1} n^k} \approx  \frac{U_n e^{\sigma t}}{qs}  \left( \frac{1}{\sigma n^2} - \frac{2}{\sigma^2 n^3}+ \frac{6}{\sigma^3 n^4}... \right).   \]
As expected, for $\sigma \rightarrow \infty$ or $n \rightarrow \infty$ we recover the characteristic $n^{-2}$ decay of an exponentially expanding population. Keeping the leading order term, we obtain
\[ F(n,t) =  \frac{U_n e^{\sigma t}}{qs \sigma n^2} \Rightarrow f(x) dx = \frac{U_n dx}{\sigma x^2}.\]

%%%%%%%%%%%%%%%%%%%%%%%%%%%%%%%%%%%%%%%%%%%%%%%%%%%%%%%%%
%%%%%%%%%%%%%%%%%%% APPENDIX F %%%%%%%%%%%%%%%%%%%%%%%%%%%%%%%%
%%%%%%%%%%%%%%%%%%%%%%%%%%%%%%%%%%%%%%%%%%%%%%%%%%%%%%%%%

\section{Appendix F: Implications for the structure of genealogies}

\manuallabel{genappendix}{F}

%first, the structure of geneological trees, e.g. coalescence properties of geneological trees fall out almost immediately. in the high q limit we regain the BSC. this is consistent with previous findings regarding this model
In addition to describing site frequency spectra, the stochastic jump process described by the transition probabilities allows us to probe the structures of genealogies in these populations, which gives rise to an intuitive explanation for why certain properties of genetic diversity are described by the Bolthausen-Sznitman coalescent in the limit of rapid adaptation.
Typically, structures of genealogical trees are defined by the rates of coalescence between the ancestral lineages of individuals sampled from the present. While we are ultimately interested in such coalescence rates, it is instructive to instead consider rates of coalescence in a fitness class coalescent framework. That is, given a sample of $p$ individuals from a \textit{single} fitness class, we are interested in the following question: if we follow the ancestry of these $p$ individuals by reversing backwards in time, will all $p$ individuals coalesce within the given class, or will their genealogies trace backwards through previous fitness classes before coalescing?

Note that because of the rapid exponential expansion exhibited by super-fit clones at the distribution's nose, coalescence rates in these populations exhibit a separation of timescales. Specifically, given $p$ independently sampled individuals in the mean class, it takes a time of $\mathcal{O}(q \langle \tau_q \rangle)$ backwards into the ancestral history of the sample before the fitness class that they inhabit occupies the population's high-fitness nose. Because of the class's exponential expansion up to this timepoint, coalescence before this time is rare for small $p$.

Once the ancestors of the $p$ sampled individuals occupy the nose, they may either coalesce within the current class, or their ancestry may span multiple fitness classes. An equivalent and more intuitive way of understanding this process is as follows: if all $p$ individuals are clonal with regards to beneficial mutations, then they are all derived from one successful mutant introduced from the previous class. In this case, they will all coalesce within the current class. If the $p$ individuals are not clonal with respect to beneficial mutations, then they are descended from two or more successful mutants from the previous class, and as such much coalesce at earliest within the previous class.

This formalism was studied carefully in the work of \citet{DWF} and used to derive a number of coalescent statistics and properties of genetic diversity in rapidly adapting populations. In what follows, we demonstrate that coalescence rates of individuals from one fitness class to another --- in other words, coalescence rates of high fitness mutants at the distribution's nose --- are clearly seen to obey a genealogy described by the Bolthausen-Sznitman coalescent in the limit of rapid adaptation. By deriving coalescence rates of individuals between classes, we have derived the coalescence rate of high-fitness mutants over a timescale $\langle \tau_q \rangle$.

To derive $\lambda_p$, the rates of coalescence of $p$ individuals in a single fitness class (given that this class is already near the nose), we reiterate that if $p$ individuals coalesce within a particular fitness class $k$, then they were all derived from a single mutant originating from the previous class $k-1$. Thus, $\lambda_p$ is closely related to the frequency spectrum of haplotypes in class $k$, $f_1(x)$. Equivalently, $f_1(x)$ is the density of sites at frequencies between $x$ and $x+dx$ that were introduced from a beneficial mutation from class $k-1$ into class $k$. Given $f_1(x)$, the coalescence rate of $p$ individuals in the nose is simply
\[\lambda_p = \int_0^1 \underbrace{f_1(x)}_{(a)}\underbrace{x^p}_{(b)} dx, \]
since $(a)$ gives the expected number of haplotypes with frequencies between $x$ and $x+dx$, and $(b)$ is the probability that, given such a haplotype, all $p$ individuals belong to it.

In the section on the site frequency spectrum of beneficial mutations, we have derived that
\[ f_1(x) = \frac{\sin(\pi \alpha) (1-x)^{\alpha-1}}{\pi x^{1+\alpha}} \approx \frac{1}{qx^2} \]
in the limit $q \rightarrow \infty$. In this case, the integral for $\lambda_p$ evaluates to
\[ \lambda_p = \frac{1}{q(p-1)}. \]
These coalescence rates define a genealogy obeying the Bolthausen-Sznitman coalescent, which has been associated with a wide class of models describing adapting populations \citep{Brunet2, Brunet3, Neher2}, and is also consistent with our previous findings \citep{DWF}. Note that the above derivation gives rise to a much simpler way to view genealogies in these populations, which was first discussed in \citet{DWF}, but bears repeating. Instead of considering the descendants of all individuals in the population, we merely consider the process by which one lead class generates another lead class over a rescaled `generation time' $\langle \tau_q \rangle$. The expected offspring distribution of these high fitness mutants in this modified coalescent tree is then distributed as $1/(qx^2)$ for large $q$, which defines a genealogy obeying Bolthausen-Sznitman statistics. The genealogical tree of $p$ individuals sampled from the mean is then roughly `star-shaped' until a time of $\mathcal{O}(q \langle \tau_q \rangle)$, at which point all the ancestors of the $p$ individuals congregate in the high fitness nose. Thereafter, their genealogies are described by the coalescence rates $\lambda_p$. One consequence of this finding is that, \textit{a priori}, we expect those diversity statistics that are set by the dynamics at the nose--- statistics which include the site frequency spectra, heterozygosities of beneficial mutations, and coalescence rates of high-fitness mutants---  to approach those predicted by the Bolthausen-Sznitman coalescent in the rapid adaptation, $q \rightarrow \infty$ limit.

%%%%%%%%%%%%%%%%%%%%%%%%%%%%%%%%%%%%%%%%%%%%%%%%%%%%%%%%%
%%%%%%%%%%%%%%%%%%% APPENDIX G %%%%%%%%%%%%%%%%%%%%%%%%%%%%%%%%
%%%%%%%%%%%%%%%%%%%%%%%%%%%%%%%%%%%%%%%%%%%%%%%%%%%%%%%%%
\section{Appendix G: Details concerning forward-time Wright Fisher simulations}
\manuallabel{simulations}{G}

We validate some of our results in the text by comparing theorized predictions to simulations. Toward this end, we implemented forward-time simulations that closely resemble evolution in the Wright-Fisher model. The details of the implementation of these simulations is described in detail in our previous work \citep{Good1}.

To measure the transition probabilities and sojourn times of mutations between fitness classes, an initially clonal population is allowed to evolve for $2 (q+1) \langle \tau_q \rangle$ generations until it reaches its steady state distribution of fitnesses. At this point, a mutation is seeded in at a frequency $x = 0.5$ in each fitness class. A new class $k$ is allowed to establish, and shortly afterwords the mutation reaches some steady frequency $x_k$ in this class. The population is then allowed to evolve until a class containing $\lceil k+q \rceil$ beneficial mutations establishes. At this point, class $k$ is (roughly) at the population's mean fitness, and the frequency of the mutant in this class is recorded. This prescription certifies that frequencies of mutations attain their long-time steady values long before they are measured. Upon the establishment of class $\lceil k+q+1 \rceil$, the frequency of the mutant in class $k+1$ is recorded. Generally, when class $\lceil k+q+i \rceil$ establishes, the frequency of the mutant in class $k+i$ is recorded. In this way, a vector of transitions $\{x_k, x_{k+1},...,\}$ is generated, until $x_i= 0$ or $x_i =  1$ in the mean class for some $i$. The transition and sojourn probabilities are then collected from 60,000 such runs for each parameter set.

Implementation of our code in Python (used to obtain transition and sojourn times) and C (used to obtain site frequency spectra) are freely available upon request.

\end{appendices}

\pagestyle{plain}
\bibliographystyle{cbe}
\bibliography{bib}

\clearpage
%
%\begin{figure}[h!]
%\begin{center}
%\includegraphics[width=0.6\textwidth]{mullerplot2.png}
%\caption{Muller plot demonstrating evolutionary dynamics in the strong selection, strong mutation regime. Haplotypes are labelled according to the beneficial mutations that they carry, which are indexed by capital letters. Beneficial mutations $A,B,C,$ and $D$ all occur on the wild-type background within a short time period. However, the success of each mutation is dependent on the chance occurrence of further beneficial mutations in its lineage. As a result, mutations $C$ and $D$ are outcompeted as the lineages containing $A$ and $B$ accrue more beneficial mutations. The frequencies of the lineages containing $A$ and $B$ also fluctuate erratically as a result of this competition. \label{mullerplot2}}
%\end{center}
%\end{figure}

\end{document}